\documentclass{article}
\usepackage{graphicx} 
\usepackage{braket} 
\usepackage{xcolor}
\usepackage{amsmath}
\usepackage{mathtools}
\usepackage{subcaption}
\usepackage{float}
\usepackage{algorithm}
\usepackage{algpseudocode}
\usepackage{subcaption}
\usepackage{enumerate}
\usepackage[normalem]{ulem} 
\usepackage[a4paper, total={6.3in, 9.7in}]{geometry}
\usepackage{array}

\newenvironment{conditions}
  {\par\vspace{\abovedisplayskip}\noindent\begin{tabular}{>{$}l<{$}@{${}={}$}l}}
  {\end{tabular}\par\vspace{\belowdisplayskip}}
  
\author{Sorana Catrina$^1$ \and Alexandra Băicoianu$^2$}
\date{%
    $^1$Faculty of Mathematics and Computer Science, Transilvania University of Brașov, Eroilor, 29, Brașov, 500036, Romania\\%
    $^2$Department of Mathematics and Computer Science, Transilvania University of Brașov, Eroilor, 29, Brașov, 500036, Romania, email a.baicoianu@unitbv.ro\\[2ex]%
}
  
\title{Quantum Tunneling: From Theory to Error-Mitigated Quantum Simulation}

\begin{document}

\maketitle

\section{Abstract}
 Ever since the discussions about a possible quantum computer arised, quantum simulations have been at the forefront of possible utilities and the task of quantum simulations is one that promises quantum advantage. In recent years, simulations of large molecules through VQE or dynamics of many-body spin Hamiltonians may be possible, and even able to achieve useful results with the use of error mitigation techniques. Simulating smaller models is also important, and currently, in the NISQ (\textit{Noisy intermediate-scale quantum}) era, it is easier and less prone to errors. This current study encompasses the theoretical background and the hardware aware circuit implementation of a quantum tunneling simulation. Specifically, this study presents the theoretical background needed for such implementation and highlights the main steps of development. Building on classic approaches of quantum tunneling simulations, this study improves the result of such simulations by employing error mitigation techniques (ZNE and REM) and uses them in conjunction with multiprogramming of the quantum chip for solving the hardware under-utilization problem that arises in such contexts. Moreover, we highlight the need for hardware-aware circuit implementations and discuss these considerations in detail to give an end-to-end workflow overview of quantum simulations. 
 
 Keywords: quantum simulation, error mitigation, multiprogramming, compiler/transpiler
 
\section{Introduction}
Simulations of natural phenomenon have always been of great interest to scientists, and ever since Feynman proposed that quantum mechanics can not be efficiently simulated on a classical computer, quantum computers have been employed for such tasks \cite{Feynman_1982, Lloyd1996UniversalQS}. 
\par Current simulations on quantum computers are, however, prone to errors. Being in the NISQ era (Near Intermediate-Scale Quantum), there are different options to deal with these impediments \cite{Preskill_2018}. Namely, one can try to reduce the noise in the system through hardware improvements and calibrations \cite{Calibration_Liu} or design algorithms that leverage the structure of noise in order to extract the useful information represented usually by the estimated value at zero noise level. This can be done either through classical post-processing techniques such as ZNE \cite{Giurgica_Tiron_2020, ZNE_2017} or REM \cite{electronics11192983, Yang_2022}, or even by employing machine learning algorithms \cite{liao2023machine}. 
\par Regarding the fields that have the possibility to contribute greatly from quantum supremacy, quantum simulations for quantum chemistry is one such example  \cite{Cirac2012GoalsAO, Daley2022PracticalQA}. The study of different materials through quantum simulations can also prove to be an important field that would benefit from quantum advantage \cite{zhu2024localization}. Implementations of such tasks that take into consideration current limitations of quantum computers already exist, and range from quantum simulations of small molecules through variational quantum eigensolvers to modelling of open systems \cite{Greenaway_2024, K_hn_2019, Lepp_kangas_2023, McCaskey_Parks_Jakowski_Moore_Morris_Humble_Pooser_2019}.
\par The current article focuses on an example of a quantum phenomenon known as quantum tunneling. It aims at simulating this on a quantum computer, the circuit of which serves as an effective tool for pushing the boundaries of existing hardware, given the potential to increase circuit depth through simulating over longer periods of time. Quantum simulations of tunneling have predominantly concentrated on a single approach to executing the given task \cite{Abouelela2020,  hegade2021experimental, Sornborger_2012}. With a focus on hardware run considerations for superconducting architectures, the study at hand aims to clarify various circuit implementation alternatives. We believe it is crucial to emphasize the compiler's significance in relation to the utilized abstract circuit for this task. Moreover, error mitigation techniques that proved efficient are discussed. Multiprogramming, a technique aimed at addressing the hardware under-utilization problem,  is also employed along with the EM techniques chosen to complete the proposed end-to-end workflow. Results of these experiments are also presented, showing an expectation value off only by $0.006$ for the transmission probability of the particle through the potential barrier after error mitigation.
\section{Theoretical Background}

To simulate Quantum Tunneling effectively, one must consider the theoretical implications inherent in such an undertaking. Therefore, this section will focus on the physics background needed for understanding this effect, namely, the Schr\"{o}dinger equation. The evolution of the wavefunction of a quantum mechanical system is described by the Schr\"{o}dinger equation: 
\begin{equation}
i\hbar \frac{\partial}{\partial t} \psi(x, t) =(-\frac{\hbar}{2m}\vec \nabla^2+V(x,t)) \psi(x, t)
\end{equation}
where
\begin{conditions}
  \hbar & reduced Planck constant \\
  \psi(x,t) & the wavefunction representing the system\\
  \vec \nabla^2 & the Laplace operator \\
  V(x,t) & the potential that represents the environment at time t  
\end{conditions}

It is evident that this constitutes a linear partial differential equation. The wavefunction, $\psi(x,t)$, assigns a complex number to each point in space, $x$, at each point in time, $t$. The potential energy term (in the equation denoted with $V$ corresponding to the environment in which the particle exists) and the kinetic energy operator form the Hamiltonian operator, which corresponds to the quantum system's total energy. For a one-dimensional system, the Laplace operator takes the following form:
$$
\vec \nabla^2 = \frac{\partial}{\partial x^2}
$$
We should note that based on the potential energy, the eigenspectrum of the wave function takes different forms. In this study, we will consider a square-well potential, which we will present in Section \ref{sec-quant-tunneling} 


The mathematical formulation of quantum mechanics states that the wavefunction, previously denoted as $\psi(x,t)$ is a statevector, $\ket{\psi}$, belonging to a Hilbert Space, $\mathcal{H}$. More information on the formalism of quantum mechanics can be found in Griffith's Introduction to Quantum Mechanics \cite{Griffiths_Schroeter_2018}. In Dirac notation, the time-dependent Schr\"{o}dinger equation takes the following form:
\begin{equation}
    i\hbar \frac{\partial}{\partial t} \ket{\psi(x, t)} =\hat{H} \ket{\psi(x, t)}
\end{equation}
\begin{equation}
    \hat{H}  = \hat{V} + \hat{K}    
\end{equation}
where
\begin{conditions}
  \ket{\psi(x,t)} & the statevector of the quantum system  \\
  \hat{H} & the Hamiltonian operator \\
  \hat{V} & the potential operator \\
  \hat{K} & the kinetic operator
\end{conditions}
Here,  $\hat{H}$ represents the \textit{Hamiltonian operator} and as specified earlier, this is the sum between the potential energy operator and the kinetic energy operator. $\hat{H}$ is a hermitian operator whose eigenvalues represent the system's energy levels. Following the separation of variables, the time-independent Schr\"{o}dinger equation along with the time evolution of an initial state $\ket{\psi(0)}$ (which can be described by the \textit{time evolution operator}, $\hat{U}$) are presented bellow. We can see that Equation \ref{eq-tischr} is an eigenvalue equation, therefore wavefunction is an eigenfunction of the \textit{Hamiltonian operator} with corresponding eigenvalues E.

\begin{equation} \label{eq-tischr}
    \hat{H} \ket{\psi(x, t)} = E \ket{\psi(x, t)}    
\end{equation}
\begin{equation} \label{eq-time-evolution}
\ket{\psi(x,t)} = \hat{U} \ket{\psi(x,0)} = e^{-i\hat{H} t} \ket{\psi(x, 0)}    
\end{equation}
where
\begin{conditions}
  \ket{\psi(x,t)} & the statevector of the quantum system  \\
  \hat{H} & the Hamiltonian operator \\
  E & the energy of the system \\
  \hat{U} & the time evolution operator
\end{conditions}

\subsection{Quantum Tunneling} 
\label{sec-quant-tunneling}
Quantum tunneling is a phenomenon that is unique to quantum mechanics, it defies explanation through classical mechanics models, and it shaped our understanding of the world around us. Its significance extends across various fields, such as Scanning Tunneling Microscopy \cite{PhysRevB_scanning_tunneling_micro} and Josephson Junctions, which are pivotal in the development of superconducting quantum computers \cite{Huang_2020}. Another example is quantum dots that also harness the principles of quantum tunneling and are used in other quantum computer architectures, such as spin qubit quantum computers, first proposed in \cite{QuantumDots_Divicenzo}. This underscores the critical necessity to delve into the study of this effect, given its far-reaching implications across numerous fields.
\par The effect of Tunneling happens between two wells separated by a potential barrier. That is, the potential takes the form presented in Equation \ref{eq-quantum-tun}, where \textit{a} is the size of the potential barrier. Solving the Schr\"{o}dinger equation for the given potential and a kinetic energy gives rise to a wavefunction that tunnels between the wells. This phenomenon will be visible in the simulations presented starting from Section \ref{sec:four-qub-quant-sim}. 

\begin{equation} \label{eq-quantum-tun}
    V(x) = \begin{cases}
    0, x<0 \\
    V_0, 0 \leq x \leq a \\
    0, x > 0
            \end{cases} \\
    a = \text{size of potential barrier}
\end{equation}

\section{Implementation considerations}

\par To successfully simulate quantum tunneling in a quantum computing environment, we need to implement the Hamiltonian operator and model the initial wavefunction $\ket{\psi(x,0)}$. This is crucial for executing both Equations \ref{eq-tischr} and \ref{eq-time-evolution}. To accomplish this, we must discretize both space and time. This is necessary because quantum computers operate with a fixed number of qubits, necessitating the discretization of infinite space for accurate simulation.

Furthermore, as Equation \ref{eq-time-evolution} suggests, we must simulate the system across various time intervals. This entails implementing the operator $\hat{H}t$ for each time step and sequentially applying it to the initial wavefunction. 

 In the following section, we will explore the implementation of the aforementioned equations on a quantum computer and examine various approaches as necessary. In Section \ref{sec-disc-space-and-time}, we delve into the discretization process and explore the requisite adaptations essential for correctly and efficiently simulating quantum tunneling on a quantum computer, such as the Trotter-Suzuki approximation \cite{Trotter_Suzuki_1, Trotter_Suzuki_2}.

\subsection{Discretizing time and space} 
\label{sec-disc-space-and-time}

\textit{Discretizing time} 
\par Considering the time evolution of the system, in order to simulate how the system evolves, we need to implement \textit{the time evolution operator}. Equation \ref{eq-time-evolution} can be further expanded to highlight the use of the kinetic and potential operators and how the wavefunction changes after $\Delta t$ time as in Equation \ref{eq-time-evolution-expanded}.
\begin{equation} \label{eq-time-evolution-expanded}
\vert \psi(x, t+\Delta t) \rangle = e^{-i\hat{H}\Delta t}\vert \psi(x, t) \rangle = e^{-i(\hat{K} + \hat{V})\Delta t}\vert \psi(x, t) \rangle
\end{equation}
where
\begin{conditions}
  \Delta t & the amount of time after wavefunction is left to evolve after time t\\
\end{conditions}
\par It is important to notice that if we implement the operator $e^{(\hat{K} + \hat{V}) \Delta t}$ for a small enough $\Delta t$ we would be able to simulate the time evolution of the entire system by sequentially applying this operator. However, an issue arises when attempting to utilize Equation \ref{eq-time-evolution} by decomposing into terms containing the kinetic and potential energy operators, which is highlighted in Equation \ref{eq:non-commuting}.
\begin{equation} \label{eq:non-commuting}
e^{-i\hat{H}\Delta t} = e^{-i(\hat{K} + \hat{V})\Delta t} \neq e^{-i\hat{K} \Delta t}e^{-i\hat{V}\Delta t} 
\end{equation}

\par This is because, in general, potential and kinetic energy operators do not commute. It is to remember that the potential operator is often dependent on $\hat{x}$ and that $\hat{K} = \frac{\hat{p}^2}{2m}$; taking into consideration that $[\hat{x}, \hat{p}] = i\hbar$ shows that position and momentum operators do not commute. Therefore, we will need to use \textit{Suzuki-Trotter approximation}. We will mostly use the first order formula given by Equation \ref{eq-first-order-st}, but higher order approximations can also be used \cite{Trotter_Suzuki_1}. For example, the second order Suzuki-Trotter approximation is given by Equation \ref{eq-second-order-st} \cite{Trotter_Suzuki_2}. 
\begin{equation}\label{eq-first-order-st}
e^{-i\hat{H}\Delta t} \approx e^{-i\hat{K} \Delta t}e^{-i\hat{V}\Delta t}+ O((\Delta t)^2)
\end{equation}
\begin{equation} \label{eq-second-order-st}
e^{-i\hat{H}\Delta t} \approx e^{-i\hat{V} \frac{\Delta t}{2}}e^{-i\hat{K}\Delta t}e^{-i\hat{V} \frac{\Delta t}{2}} + O((\Delta t)^3)
\end{equation}

\textit{Discretizing space}

Since the Schr\"{o}dinger equation is given in continuous space, we encounter a challenge in implementation due to the finite number of qubits available. We can therefore discretize the interval $0\leq x \leq L$ with interval spacing $\Delta l$ within the boundary region with a periodic boundary condition $\vert \psi(x+L, t) \rangle = \vert \psi(x, t) \rangle$ The wave can therefore be mapped to a $n$-qubit register as in Equation \ref{eq-discretize-space} where k represents the particle location corresponding to binary number k.
\begin{equation} \label{eq-discretize-space}
    \vert \psi(x+L, t) \rangle = \sum_{k=0}^{2^n-1}\psi(x_k, t) \vert k \rangle
\end{equation}

\subsection{Implementing the operators}

In accordance with the Suzuki-Trotter approximations, to simulate quantum tunneling, the following operators need to be implemented: 

\begin{enumerate}
    \item The Kinetic Energy Operator, $e^{\hat{K}}$ 
    \item The Potential Energy Operator, $e^{\hat{V}}$
\end{enumerate}

\subsubsection{The Kinetic Energy Operator} \label{sec:kinetic-energy-operator}
The kinetic energy operator can be represented with the help of the momentum operator as shown in Equation \ref{eq-kintic-energy-operator}, where $\hat{p}$ is the momentum operator.
\begin{equation} \label{eq-kintic-energy-operator}
 \hat{K} = \frac {\hat{p}^2}{2m}.   
\end{equation}
where
\begin{conditions}
  \hat{p} & the momentum operator \\
  m & the mass of the particle 
\end{conditions}
Different from coordinate space, in momentum space $\hat{p}$ is diagonal. The potential operator, however, is dependent on $\hat{x}$, which is diagonal in coordinate space. Consequently, it is advantageous to switch from coordinate space to momentum space to apply the kinetic energy operator and come back to coordinate space to apply the potential operator (this is because the construction of diagonal operators is easier). In quantum mechanics, switching from coordinate space to momentum space is done with the help of the Fourier Transform, as shown in Equation \ref{eq-fourier-direct-transform}. The inverse transformation is done through Equation \ref{eq-inverse-fourier-transform}. 

\begin{equation} \label{eq-fourier-direct-transform}
\tilde{\psi}(p, t) = \frac {1} {2\pi \hbar} \int_{-\infty}^{\infty}dx\psi(x, t)*e^{-\frac {-ipx}{\hbar}}
\end{equation}
\begin{equation} \label{eq-inverse-fourier-transform}
\psi(x, t) = \frac {1} {2\pi \hbar} \int_{-\infty}^{\infty}dp\tilde{\psi}(p, t)*e^{-\frac {-ipx}{\hbar}}
\end{equation}

Considering the same limitations, we will need to use the discrete formula of the Fourier Transform, and moreover, we will use the Quantum Fourier Transform (QFT), shown below. We will use the inverse Quantum Fourier Transform (IQFT) to convert back to coordinate space. 

\begin{equation*}
    \begin{multlined}
        \vert j \rangle \mapsto \frac{1}{\sqrt{N}}\sum_{k=0}^{N-1}\omega_N^{jk} \vert k \rangle \\
\omega_N^{jk} = e^{2\pi i \frac{jk}{N}}
    \end{multlined}
\end{equation*}

As mentioned earlier, in momentum space, the kinetic operator is diagonal. Using the notation in Equation \ref{eq-not-k}, the operator $e^{-i\hat{k}^2\Delta t}$ can be represented with the use of the phase gate, as shown in  Equation \ref{eq-mom-representation} with the Pauli Z gate \cite{Abouelela2020, Shokri_2021}.

\begin{equation} \label{eq-not-k}
    \ket{k} = \sum_{i=0}^{n} 2^i j_{i+1}
\end{equation}
\begin{equation} \label{eq-mom-representation}
    e^{-i \hat{k}^2 \Delta t} = exp ( \frac{i \phi}{2^{2n-3}} (1+\sum_{j=1}^n 2^{j-1} \hat{Z}_j )^2)    
\end{equation}

With this formulation, we can easily see the one qubit operations. However, in order to better see the two-qubit operations, we have to open the sum. Focusing on the two-qubit terms, we get:

$$
S = \sum_i \sum_j 2\frac{2^{j-1}2^{i-1}}{2^{2n-3}}\hat{Z_i}\hat{Z_j} = 
\sum_i \sum_j \frac{2^{i+j-1}}{2^{2n-3}}\hat{Z_i}\hat{Z_j} =
\sum_i \sum_j \frac{1}{2^{2n-i-j-2}} \hat{Z_i}\hat{Z_j}
$$

Since in this notation, $i$ and $j$ start for 1 and when implementing indexing starts from 0, we need to change the start and ending point and rewrite the sum as follows:

\begin{equation}
\begin{aligned}  \label{eq-two-qubit-rot-kinetic-energt}
    i & \rightarrow i'+1 \\ 
j & \rightarrow j'+1  \\
S & = \sum_{i'} \sum_{j'} \frac{1}{2^{2n'-(i'+1)-(j'+1)-2}} \hat{Z_i}\hat{Z_j} \\
& = \sum_{i'} \sum_{j'} \frac{1}{2^{2n'-i'-j'-4}} \hat{Z_i}\hat{Z_j}
\end{aligned}
\end{equation}

\begin{algorithm}
\caption{Kinetic Energy Operator Implementation}\label{alg:kineticEnergy}
  \begin{flushleft} 
    \textbf{Input} $\Delta t$ representing the time step size needed by trotterization, $n$ representing the number of qubits, $circuit$ the quantum circuit with two registers $q$, and $aux$ as auxiliary register \\
    \textbf{Output} The updated circuit
    \end{flushleft}
\begin{algorithmic}[1]

\For{$i \in range(n)$}
    \State $circuit.p(\frac{\phi}{2^{2n-3}}, q[i])$ \Comment{The single qubit rotations from Equation \ref{eq-mom-representation}}
\EndFor
\For{$i \in range(n)$}
    \For{$j \in range(i+1, n)$}
        \State $apply\ cx\ on\ q[i],\ aux $ 
        \State $apply\ cx\ on\ q[j],\ aux $
        \State $apply\ phase\ gate\ with\ angle\ \frac{1}{2^{2n-i-j-4}}\ on\ aux $ \Comment{The two-qubit rotations}
        \State $apply\ cx\ on\ q[j],\ aux $
        \State $apply\ cx\ on\ q[i],\ aux $
    \EndFor
\EndFor
\end{algorithmic}
\end{algorithm}

\begin{figure}
    \centering
    \includegraphics[scale=0.40]{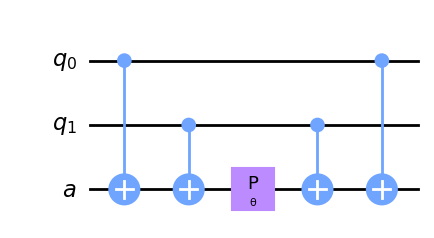}
    \caption{$Z_1 \otimes Z_2$ Hamiltonian implementation using ancilla qubits}
    \label{fig:example-hamiltonian-implementation}
\end{figure}

The implementation can be accomplished in various ways. One approach is to follow the schema using auxiliary qubits \cite{QuantumComputationAndInformation}, but one can also manage without such auxiliary qubits in simulations of 2 and 3 qubits \cite{hegade2021experimental, Sornborger_2012}. Using an auxiliary qubit is done in order to keep count of the parity of the qubits (the phase shift applied to the system is $+i \Delta t$ if the parity of the n qubits in the computational basis is even). This is highlighted in Algorithm \ref{alg:kineticEnergy} and illustrated for an example hamiltonian in Figure \ref{fig:example-hamiltonian-implementation}. The first section of the algorithm is for the single-qubit rotations, while the second is for the two-qubit terms in the above mentioned equations. We mention that lines 9-10 are meant to \textit{undo} the parity counting in order to account for reusability. 

However, we have to keep in mind that using an ancilla qubit will increase the hardware mapping depth for superconducting current architectures since the connectivity is limited. More information on implementations considering hardware constraints will be presented in Section \ref{Preparing-to-run-on-hardware}.



\subsubsection{Potential Energy Operator}
\label{Potential-energy-operator}

Implementation of the potential energy operator is straightforward in coordinate representations: it is done by applying Z rotations on the qubit (or CZ operations on different qubits), depending on what type of potential we want. It is important to point out that we will follow qiskit ordering, that is, little endian. For example, we can apply the controlled Pauli Z operator on the highest-order qubit to implement multiple wells, as shown in the equation below. Similarly, to have only one well, we can apply it to the lowest-order qubit. 
$$
e^{i\hat{V}\Delta t} = e^{-i v \sigma_z^{n-1} \Delta t} = I \otimes I \otimes I \otimes  ... \otimes \sigma_z 
$$
here, $\sigma_z^{n-1}$ is the Pauli Z operator on the $n-1$ qubit.
\subsubsection{Quantum Fourier Transform}

We will need to implement the Quantum Fourier Transform to allow switching between coordinate space and momentum space representations.
The Quantum Fourier Transform maps a quantum state $\ket{X}$ onto another quantum state $\ket{Y}$ by the following formulas:
    \begin{equation}
        \begin{split}
            \ket{X} & = \sum_{j=0}^{n-1}x_j\ket{j} \\
            \ket{Y} & = \frac{1}{\sqrt{n}} \sum_{j=0}^{n-1}y_j\ket{j} \\
            y_k &= \frac{1}{\sqrt{n}} \sum_{j=0}^{n-1} e^{2\pi i \frac{jk}{n} } x_j
        \end{split}
    \end{equation}
In order to implement the QFT, we can represent it as follows \cite{QuantumComputationAndInformation}:

\begin{equation}
    \begin{aligned}
    QFT \ket{x} &= \frac{1}{\sqrt{N}}  e^{2\pi i \frac{jk}{n} } x_j  \\
                & = \frac{1}{\sqrt{N}} (\ket{0} + e^{\frac{2\pi i}{2}x} \ket{1}) \otimes (\ket{0} + e^{\frac{2\pi i}{2^2}x} \ket{1}) \otimes ... \otimes  (\ket{0} + e^{\frac{2\pi i}{2^n}x} \ket{1})
    \end{aligned}
\end{equation}

\subsubsection{Initial State} \label{sec:initial-state}
For simulations where the potential is $e^{i\hat{V}\Delta t} = e^{-i v \sigma_z^{0} \Delta t}$ (meaning we have a well on every other site), the initial state has $\ket{1}$ where we want our state to start. For every other type of potential, a gaussian wave is used as the main form:

$$
\ket{\psi} = \frac{1}{\sqrt{2\pi}\sigma} e^{-\frac{1}{2}(\frac{x-\mu}{\sigma})^2}
$$
However, for moving waves, a momentum component is also added:

$$
\ket{\psi} = \frac{1}{\sqrt{2\pi}\sigma} e^{-\frac{1}{2}(\frac{x-\mu}{\sigma})^2} e^{-ipx}
$$
This works well for simulations with barriers since the wave is confined. A Gaussian wave is also used for free particle simulations. In QM, due to the need for normalization, the \textit{wavepacket} is represented as a superposition of such plane waves.

$$
\psi(x, t) = \frac{1}{\sqrt{2\pi}} \int_{-\infty}^{\infty}dk \tilde{\varphi}(k) e^{-\omega(k)t}e^{ikx} = \frac{1}{\sqrt{2\pi}} \int_{-\infty}^{\infty}dk \tilde{\varphi}(k) e^{ikx} 
$$

\subsection{Circuit Overview}

Having discussed implementations of the kinetic and potential operator in previous sections as well as the initial state considered, we can now provide an overview of the circuit implementation. We want to simulate on the quantum computer Equation \ref{eq-time-evolution}.  

\begin{figure}[h]
    \centering
    \includegraphics[scale=0.25]{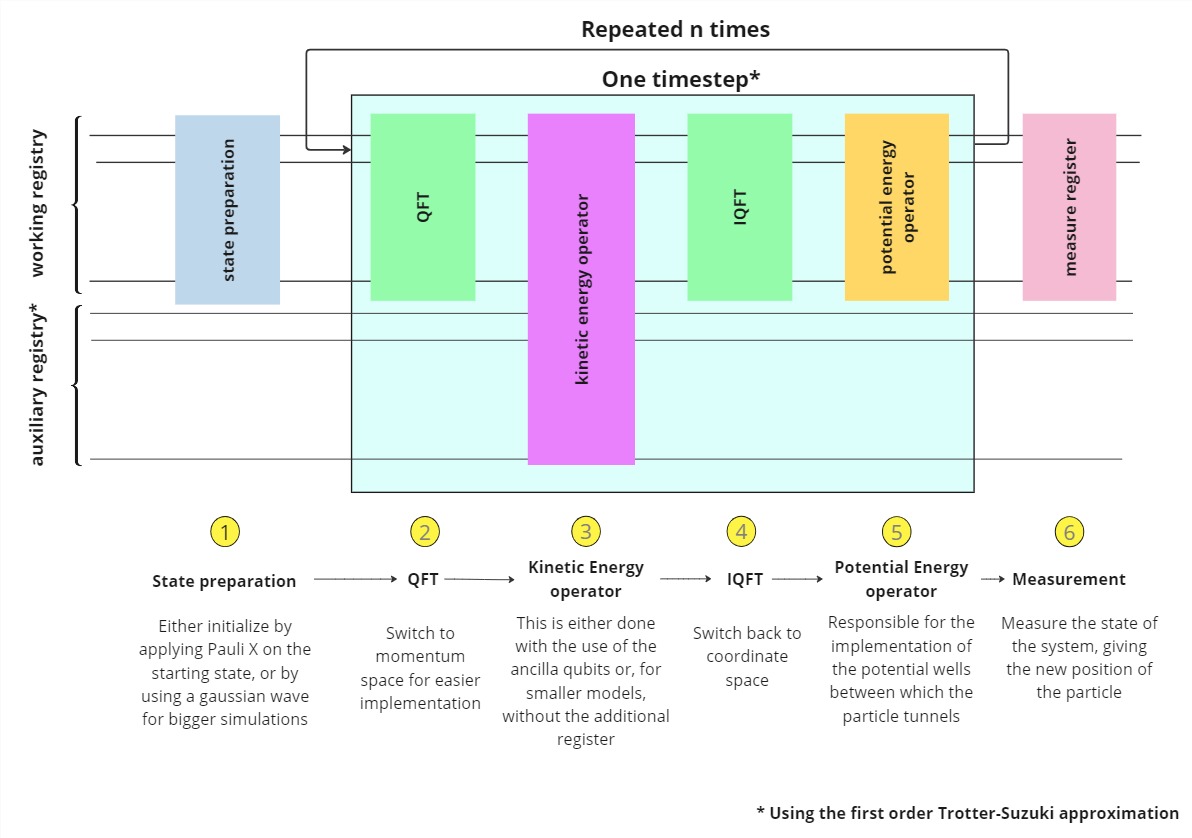}
    \caption{Full circuit overview}
    \label{fig:impl-overview}
\end{figure}

Upon considering the operators' implementation, we can provide an overview of the essential steps necessary for the creation of the circuit, as depicted in Figure \ref{fig:impl-overview} and is as follows:
\begin{enumerate}[Step 1:]
\item This is represented by state preparation. This entails preparing $\ket{\psi(x,0)}$, as discussed in Section \ref{sec:initial-state}. This state can be a Gaussian wave for systems where the space has been discretizing using more qubits. Otherwise, the wave is approximated to the state containing a flipped qubit on the position of the particle
\par After this preparation, one can start implementing the time evolution with the use of the Suzuki-Trotter Approximation. One time step is represented by steps 2-5, these implement the Kinetic and Potential energy operators, respectively.
\item The QFT is applied to the quantum circuit. As mentioned in Section \ref{sec:kinetic-energy-operator}, one will first switch to momentum space with the use of the QFT and then implement the kinetic operator. 
\item The Kinetic Energy Operator is implemented. This can be done either with or without ancilla qubits. 
\item Switching back to coordinate space is represented by the Inverse QFT.
\item Implementation of the Potential Energy Operator, as described in Section \ref{Potential-energy-operator}.
\item The last step is measuring the working register, which gives us the position of the particle.
\end{enumerate}

\begin{figure}
    \begin{subfigure}[b]{\textwidth}
        \centering
        \includegraphics[scale=0.25]{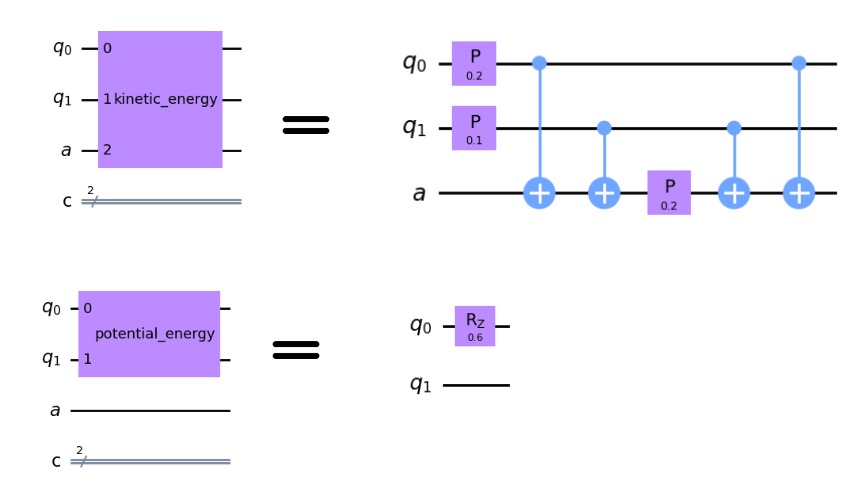}
        \caption{Implementation of operators}  
    \end{subfigure}
    \begin{subfigure}[b]{\textwidth}
        \centering
        \includegraphics[scale=0.25]{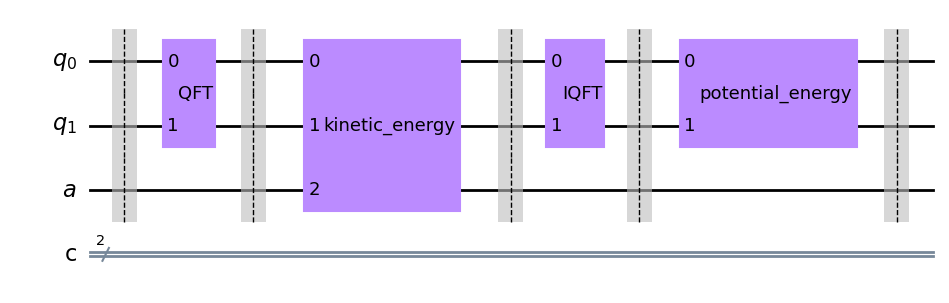}
        \caption{General one timestep}
        \label{subfig:general-time-step}
    \end{subfigure}
    \caption{High level overview of 2 qubit operations}
    \label{fig:op-implementation}
\end{figure}

For implementation, we used \textit{qiskit}, but since it is compiled to QASM the code can be reused for other languages as well. The representation of operators for a 2 qubit example, as presented in the first section, is shown in Figure \ref{fig:op-implementation}.
 A high-level view of the circuit for a 2 qubit implementation, where an ancilla qubit is used for the kinetic energy implementation, is present in Figure \ref{subfig:general-time-step} following the scheme presented in Section \ref{sec:kinetic-energy-operator} with ancilla qubits \cite{QuantumComputationAndInformation}.




\subsection{Four qubits Quantum Tunneling Simulation } \label{sec:four-qub-quant-sim}

This section presents results for \textit{4 qubit quantum tunneling simulation} in a noiseless environment. By analyzing these results, we can better understand the quantum tunneling phenomena and highlight how the parts discussed earlier come together to form a quantum tunneling simulation. Since this is a noise-free environment, the simulator employed does not account for possible errors, which are inevitable with today's quantum computers. Leveraging \textit{qiskit}, this simulator is accessible via Aer, offering users the choice between noiseless and noisy simulation modes.
\par Regarding implementation, our approach incorporated the kinetic energy using an auxiliary qubit. The potential and QFT are implemented as discussed in earlier sections.

\begin{table}[htbp]
    \centering
    \caption{Experiments Setup for 4 qubit quantum tunneling simulation}
    \label{tab:results-sim-4q}
    
    \begin{center}
    \begin{tabular}{||c c c c c||}  
         \hline
         Type of potential & Initial wave & $\Delta t$ & Number of time steps & Potential type  \\ [0.5ex] 
         \hline\hline
         Wall Potential & Gaussian wave centered at $\ket{0100}$ 
 & 0.1 & 20 & 1xxx  \\ 
         Two wells &  Gaussian wave centered at $\ket{0100}$ & 0.1 & 40 & x11x  \\
         Multiple wells & $\ket{1000}$ & 0.1 & 20 & xxx1 \\
         \hline
    \end{tabular}
    \end{center}
    
\end{table}

\begin{figure}[h]
    \centering
    \begin{subfigure}[b]{0.3\textwidth}
        \centering
        \includegraphics[scale=0.12]{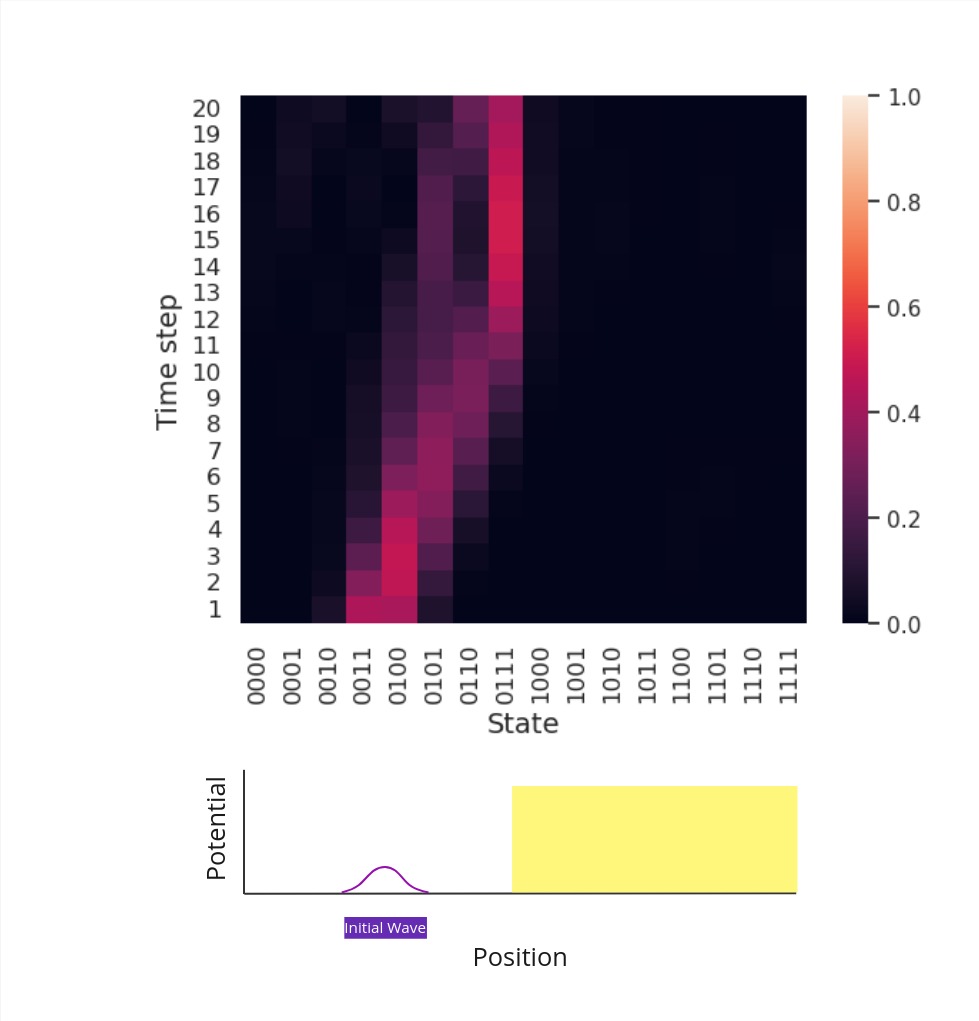}
        \caption{Simulation of wall potential}
        \label{subfig:wall-potential}
    \end{subfigure}
    \begin{subfigure}[b]{0.3\textwidth}
        \centering
        \includegraphics[scale=0.12]{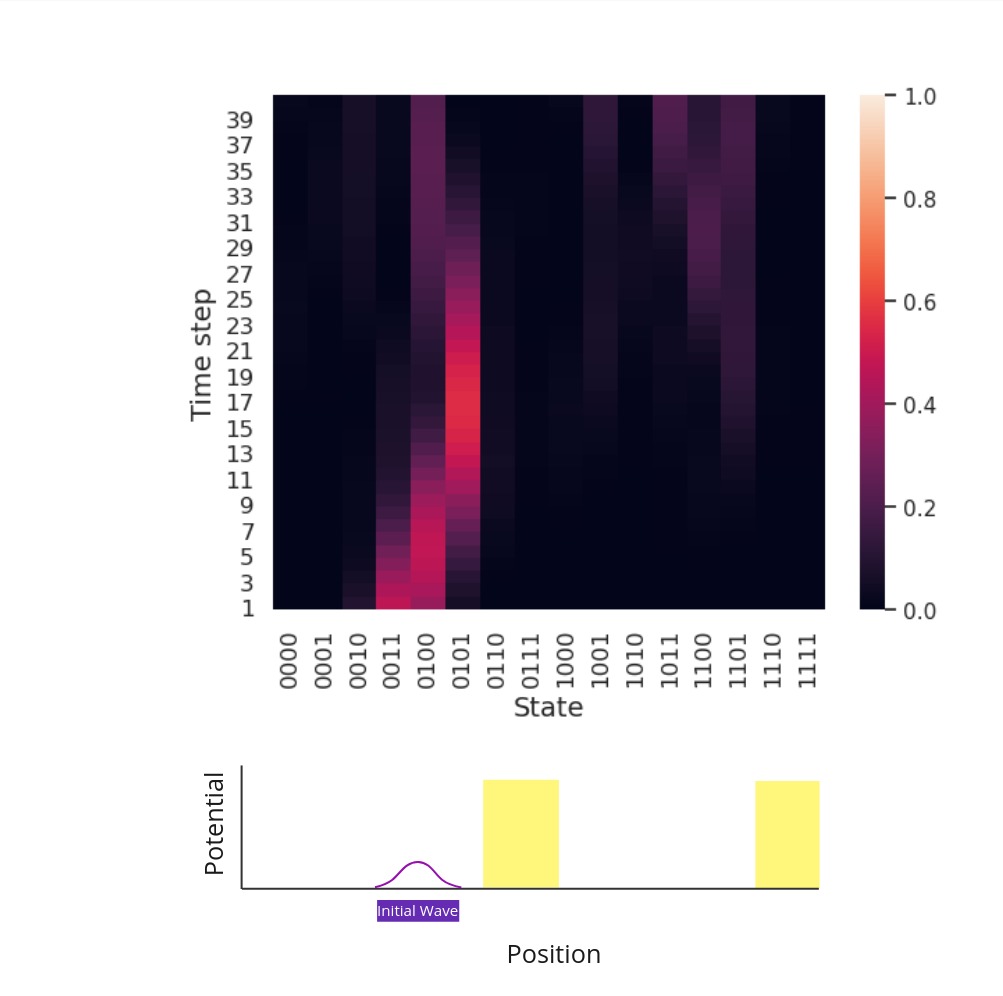}
        \caption{Two wells}
        \label{subfig:two-wells-potential}
    \end{subfigure}
    \begin{subfigure}[b]{0.3\textwidth}
        \centering
        \includegraphics[scale=0.12]{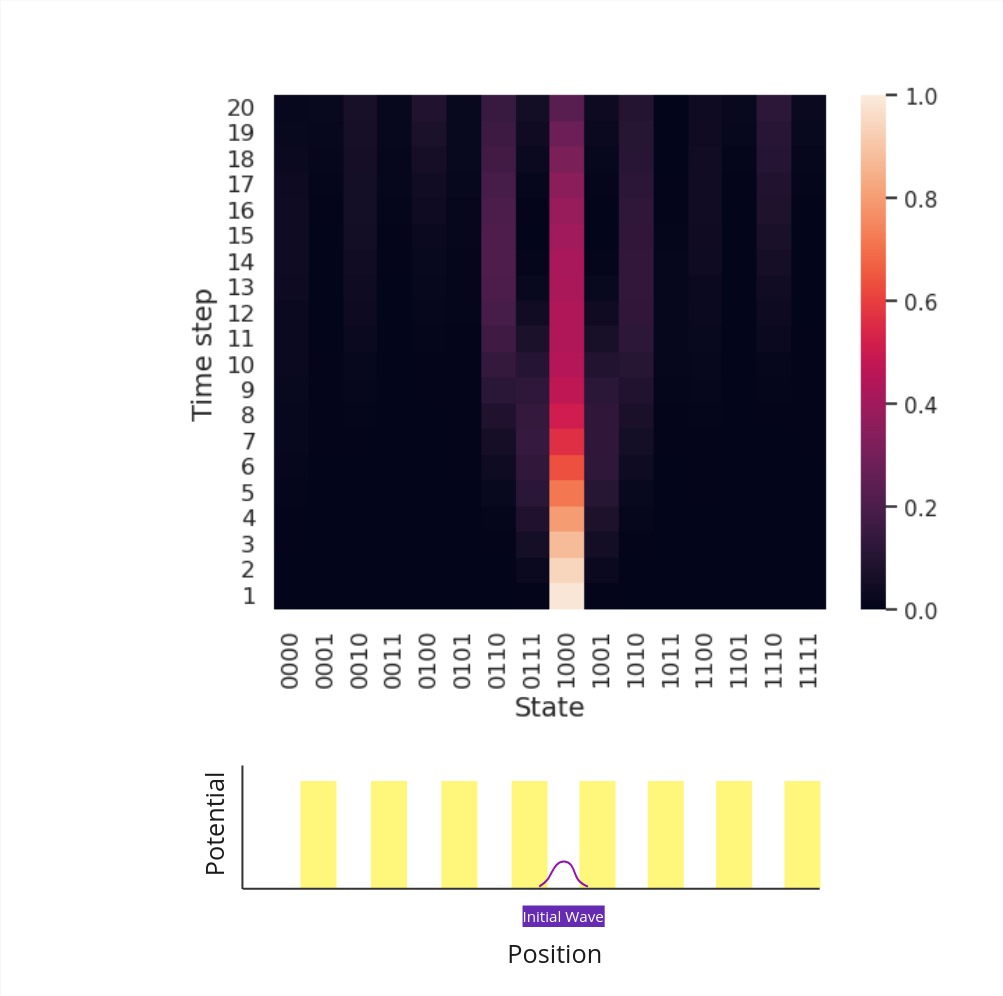}
        \caption{Multiple wells}
        \label{subfig:multiple-wells-potential}
    \end{subfigure}
    \caption{Each diagram reperesents a 4q noiseless simulation. The bottom image presents the profile of the potential, along with the initial position of the wave that can also be observed at timestep 0 of the diagram}
    \label{fig:wells}
\end{figure}

The setup of the experiments is summarised in Table \ref{tab:results-sim-4q}. Potential type refers to positions in our simulation that have a non-null potential attributed to them in the potential landscape. The x matches for either 0 or 1. For example, for the two well simulation which has a potential type of "x11x", we can see in Figure \ref{subfig:two-wells-potential} that there is non zero potential (represented by the yellow box) at positions: 0110, 0111, 1110, 1111. 
\par
\begin{figure}
    \centering
\includegraphics[scale=0.10]{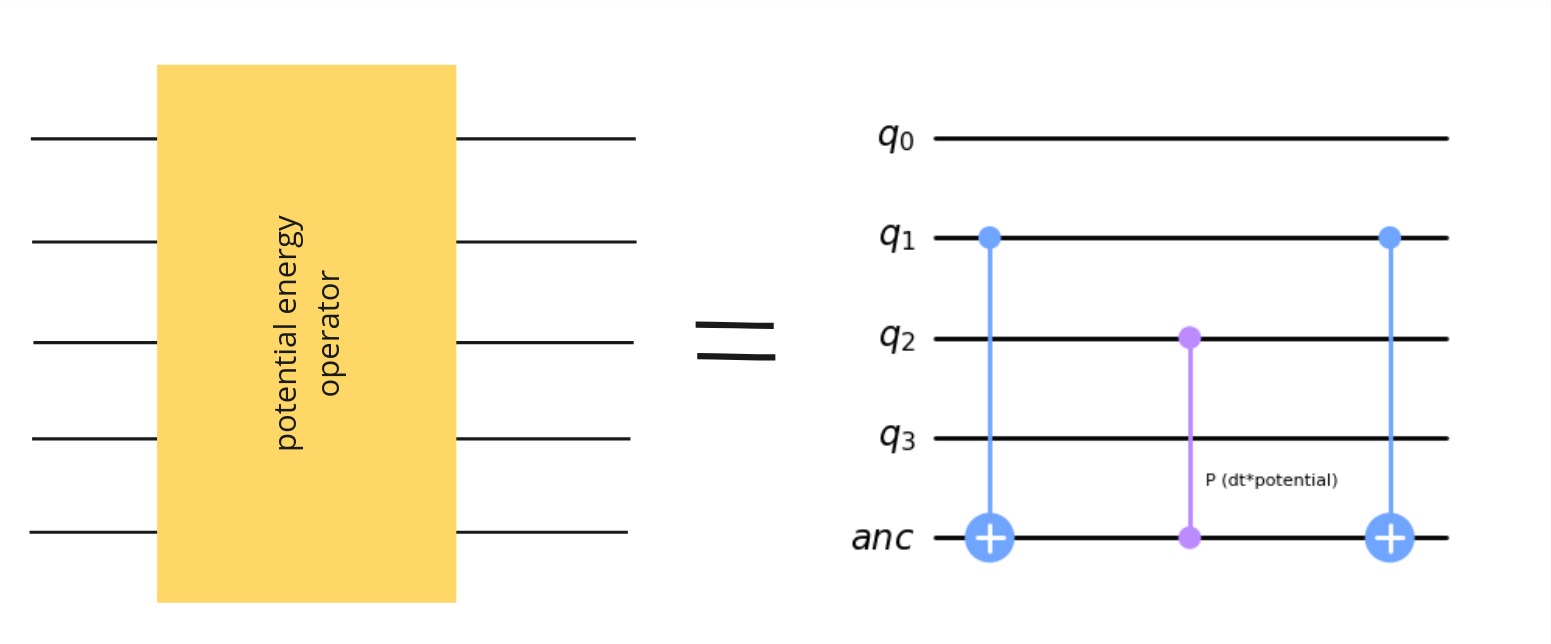}
    \caption{Implementation of x11x type potential}
    \label{fig:x11x-potential-implementation}
\end{figure}

 The wall and multiple wells potential are discussed in Section \ref{Potential-energy-operator}. Potential type x11x can be achieved through a filter-type gate, as suggested in \cite{Abouelela2020}. This should be rather intuitive, since we need to apply a phase corresponding to this operator to the states that have a non-zero potential in our modelling. This can be achieved through controlled gates using the auxiliary qubit. It can be observed that the ancilla qubit is convenient in this type of simulation since it can be used for both the potential operator and the kinetic energy operator. The implementation of the x11x barrier is presented in Figure \ref{fig:x11x-potential-implementation}.
The results are presented in Figure \ref{fig:wells}, and for \ref{subfig:two-wells-potential} and \ref{subfig:multiple-wells-potential} tunneling is observed between the wells of potential. We also point out that in Figure \ref{subfig:two-wells-potential} we can see that there is less tunneling between the wells in the first timesteps, since the potential barrier is also larger. 
Running the simulation with the circuit discussed for different timesteps, one manages to gather the results presented in Figure \ref{fig:wells}. \par The first experiment, in which there is a potential wall (meaning no wells are created) reveals no tunneling happening, as expected. The results can be seen in Figure \ref{subfig:wall-potential}. We highlight that the probability of the particle being at a certain position (represented on the x-axis) is visually illustrated through the intensity of color in the diagram. When one sets up multiple wells, as in Figure \ref{subfig:two-wells-potential} and Figure \ref{subfig:multiple-wells-potential}, tunneling can be observed. By this, we mean that the particle is able to "tunnel" from one well to the other, even though there is a potential barrier separating the two. 
 \par In Figure \ref{subfig:two-wells-potential}, one can see a particle traveling to the right and after reaching the potential barrier, "tunneling" through the other well. The probability of transmission through the potential barrier is, of course, dependent on the potential. The transmission wave and the reflected wave are also observable. In Figure \ref{subfig:multiple-wells-potential}, multiple wells have been implemented, and the simulation results are illustrated. The tunneling of the particle can be observed in this image as well. We mention that the starting position of the particle is represented by the statevector $\ket{1000}$.
\par In this section, we gathered a better understanding of what a noiseless quantum tunneling simulation reveals, studying different potentials and analyzing the results within the context of a 4 qubit simulation. Nonetheless, upon transitioning to actual hardware, numerous considerations arise that necessitate attention, thereby warranting adjustments before achieving satisfactory results. These considerations will be thoroughly discussed in the next section.

\section{Running on hardware} \label{Preparing-to-run-on-hardware}

\subsection{Introduction}

\par When running on hardware, multiple factors need to be considered. We are still in the NISQ era, and therefore we need to make our circuit as prone to error as possible, this includes everything from modifying the circuit, to choosing the right layout and routing techniques.

The first thing one has to do is choose a backend. Part of the experiments were done on one of the three 7-qubit quantum computers part of the Falcon r5.11H chip and three 128-qubit quantum computers using the Eagle r3 chip. Simulations were run on the 7-qubit quantum computer Nairobi, Jakarta, and the 128-qubit quantum computer Osaka. Apart from IBM's superconducting quantum computers, others could've also been considered, such as IQM's trapped ions quantum computers. Certain advantages and disadvantages could be analyzed for each architecture. 

Once a quantum computer is chosen, we have access to the basis gates of such computer (for our case, ECR, ID, RZ, SX, X) and the chip layout (in the case of superconducting circuits).


The \textit{layout} is important since the connectivity in superconducting quantum computers is not all-to-all. For example, when two \textit{virtual qubits} which partake in different 2-qubit gates are mapped to \textit{physical qubits} that are not connected, this might end up being less efficient since one of them will need to be swapped next to the other one (therefore adding 3 two-qubit gates to the circuit that introduce noise). 
\par Also, since our circuit most of the time is not composed only of basis gates, but of composed gates as well, we need to transform our circuit to contain only allowed gates. This operation is done by the \textit{transpiler}, which has a similar role to other languages' compiler. The transpiler works by applying multiple \textit{passes} to the circuit, each doing different operations meant to optimize, layout, reduce depth and transform the circuit to run on the backend. However, we have to keep in mind that the transpiler does not always give the best option out there, we imply that two identical circuits, yielding the same statevector, may exhibit distinct implementations leading to variations in circuit depths.

\begin{figure}
    \centering
    \begin{subfigure}[b]{\textwidth}
        \centering
        \includegraphics[scale=0.35]{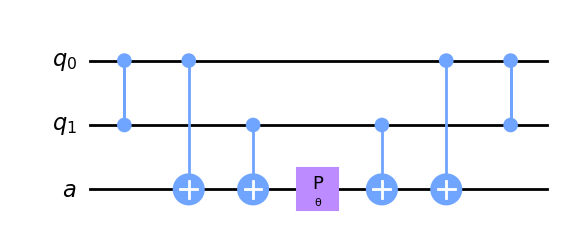}
        \caption{The circuit with virtual qubits}
        \label{subfig:circuit}
    \end{subfigure}
    \begin{subfigure}[b]{\textwidth}
        \centering
        \includegraphics[scale=0.25]{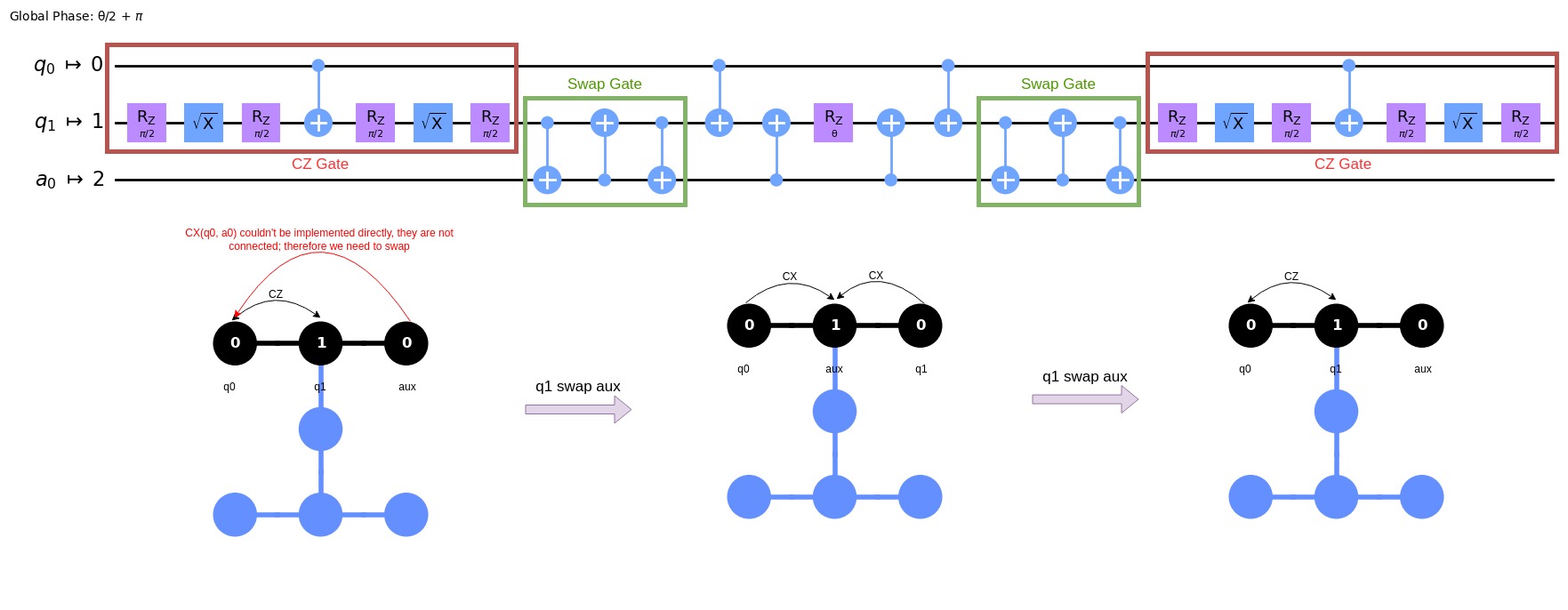}
        \caption{The transpiled circuit}
        \label{subfig:transpiled}
    \end{subfigure}
    \caption{Transpilation process}
    \label{fig:fig_transpile}
\end{figure}

In Figure \ref{fig:fig_transpile}, in the abstract circuit section, we can observe two circuits that produce, in the end, the same statevector. However, we can observe that the implementation is different, and, therefore, depending on the backend chosen, we can see how the layout can influence circuit depth. For example, on ibm\_auckland we can see that the transpiled circuit of the second implementation has a lower depth than that of the first one. Reducing 2-qubit gates plays an important role in ensuring the circuit is error-prone. However, on ibm\_nairobi, the layout of the transpiled first circuit has a lower depth. This observation, that choosing the hardware-aware implementation of a certain problem is important when it comes to superconducting circuits must be considered when designing algorithms to be run on superconducting architectures. One can also see that the depth is increased by using several swap gates that were required to map the circuit to the layout. This process, responsible for adding swap gates to ensure connectivity adherence, constitutes the \textit{routing} pass.

\begin{figure}
    \centering
    \begin{subfigure}[b]{0.4\textwidth}
        \centering
        \includegraphics[scale=0.2]{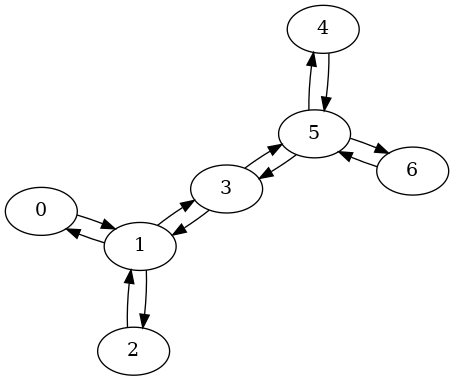}
        \caption{Coupling map for Nairobi}
        \label{subfig:nairobi}
    \end{subfigure}
    \begin{subfigure}[b]{0.4\textwidth}
        \centering
        \includegraphics[scale=0.05]{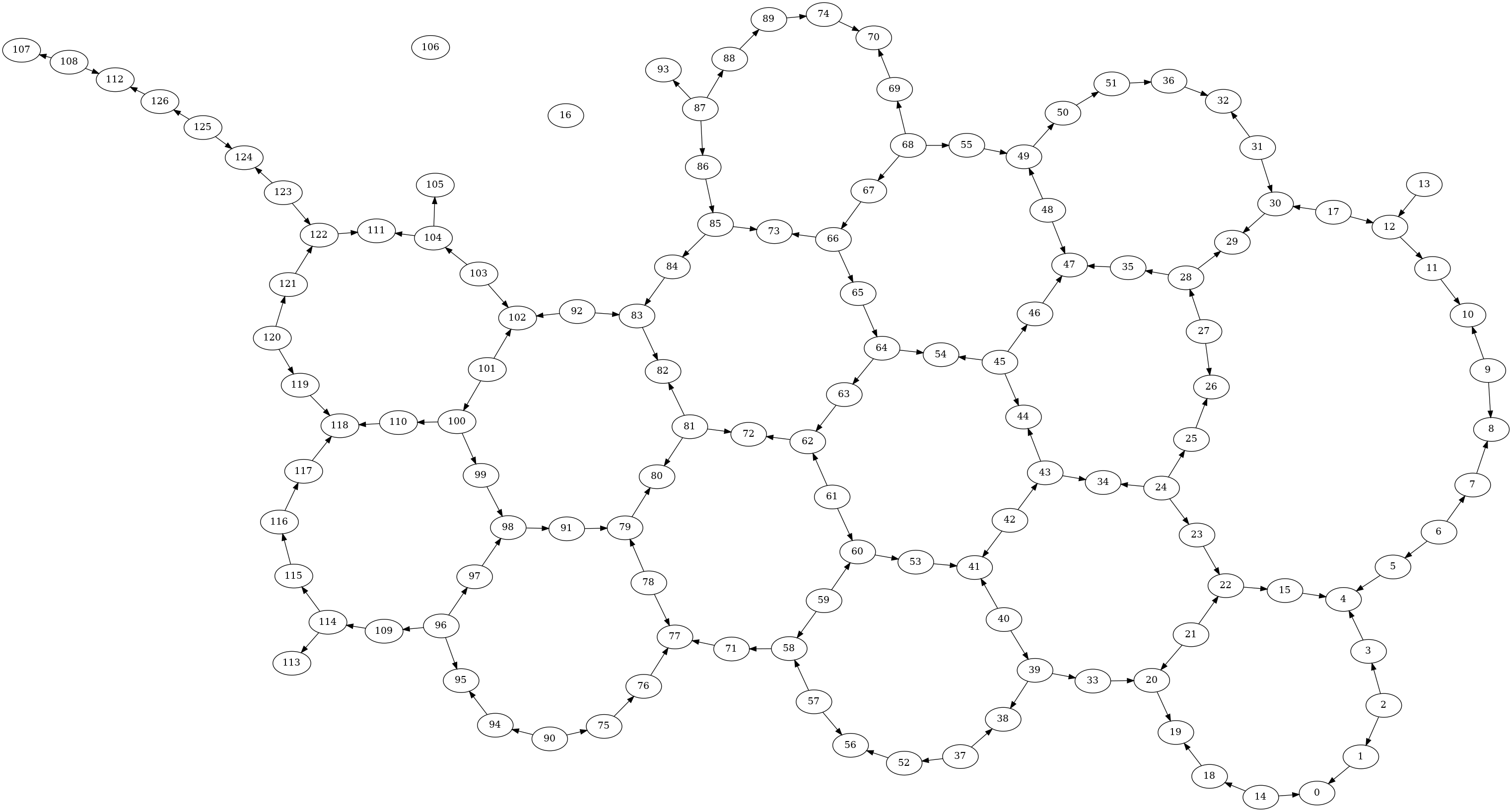}
        \caption{Coupling map for Osaka qc}
        \label{subfig:osaka}
    \end{subfigure}
    \caption{Coupling maps}
    \label{fig:coupling-map}
\end{figure}

Apart from the layout, one has to consider the coupling map as well. The coupling map was symmetric for earlier chips, such as the Falcon r5.11H. This means that we could apply controlled 2-qubit gates regardless of which physical qubit was chosen as the control or target. However, on the newer chips (Eagle r3) the connectivity is not symmetric. Illustrations for coupling maps for the Falcon chip (more specifically, the Nairobi quantum computer) and the Eagle chip (the Osaka quantum computer) are presented in Figure \ref{fig:coupling-map}. This means that, for example, the swap gate is implemented in a different way, not only because of the different basis gate (on the Eagle r3) but also due to the coupling map. We also remind that the ECR gate (part of the basis set on the Eagle r3 chip) is an echoed-cross resonance gate, equivalent to CX gate up to single qubit pre-rotations.


\subsection{Transpiler} \label{Transpilation-passes}

The transpiler is the main component that is used in order to run an abstract quantum circuit on a real quantum computer, as we've discussed earlier. This component is sometimes named compiler, depending on the programming language used. Therefore, we summarise the main components, as implemented by \textit{qiskit}, the library we chose for implementation:
\begin{itemize}
    \item Decomposing the custom gates into one and 2-qubit gates - one step here is unfolding; for our quantum simulation circuit, this means unfolding the kinetic energy operator and the potential energy operator
    \item Layout - finding the best physical qubits on which our virtual qubits will be mapped to. Another tool we used for this pass was \textit{mapomatic} for a more detailed look into the layout scoring process. 
    \item Routing - inserting swap gate in order to respect the connectivity 
    \item Translation (into basis gates - the native gates for the specific backend)
    \item Optimization - removing redundancies in the circuit, applying circuit identities in order to decrease the circuit depth. For this pass, we also used the tket compiler.
    \item Scheduling - this is an important pass for pulse-level control; but for this pass, we can also use different techniques (Dynamical Decoupling is a pass that needs scheduling afterward)
\end{itemize}

\begin{figure}
    \centering
    \includegraphics[scale=0.2]{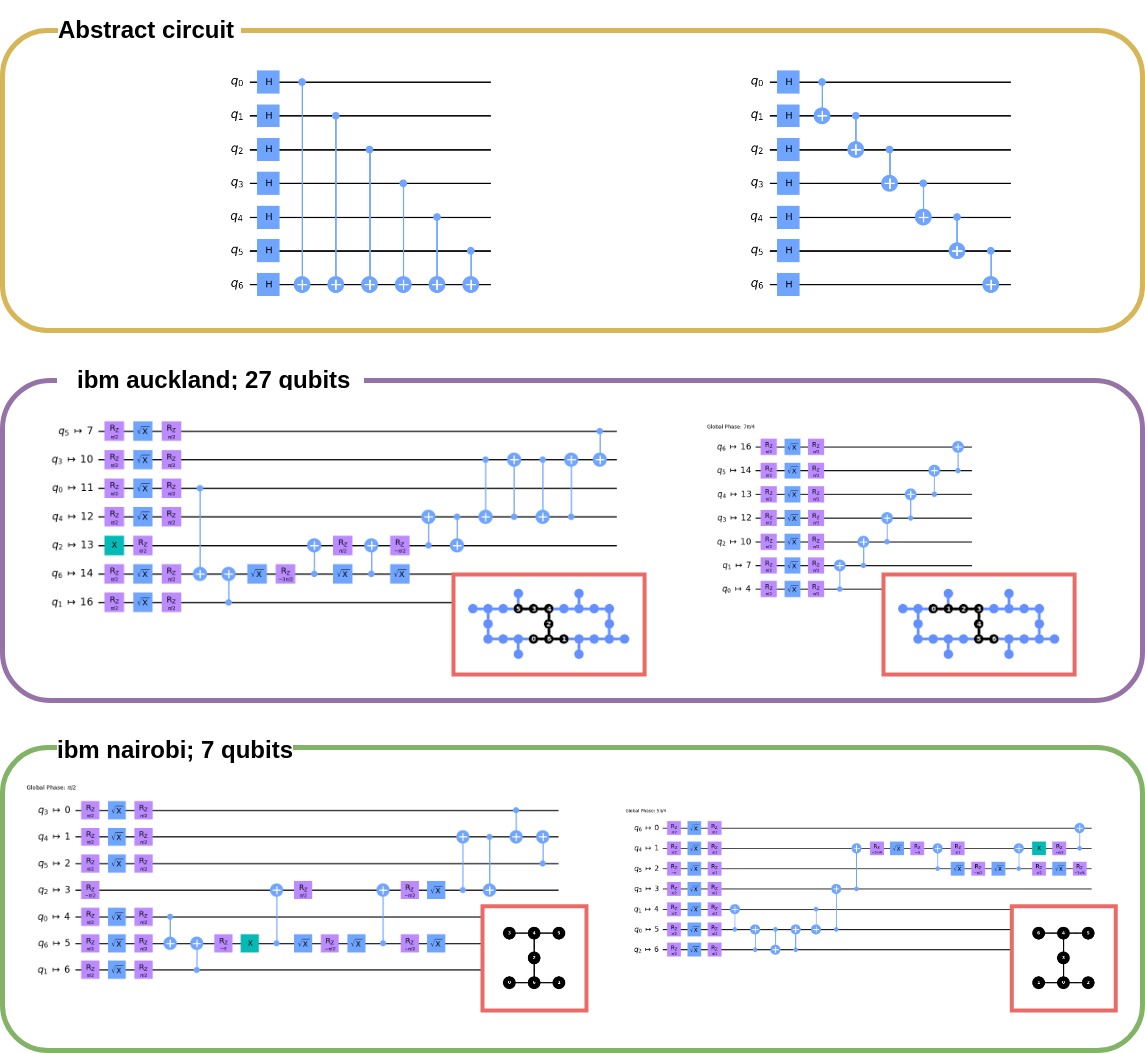}
    \caption{Illustration of why circuit design and backend influence the depth of the circuit. This calls for hardware-aware design}
    \label{fig:hardware-aware-circuit}
\end{figure}

Qiskit's transpiler is not the only one that exists and is helpful. For example, apart from this transpiler we also used tket compiler that improved the circuit depth. This is due to different passes being implemented by tket (such as the FullPeepholeOptimise pass) that when applied to our circuit, proved efficient in making it shallower. 
As mentioned earlier, one should also consider that two circuits which in the end produce the same statevector, depending on the implementation and layout of the circuit, different final circuits with different depths would result. Figure \ref{fig:hardware-aware-circuit} presents one example that highlights this idea.

\subsection{Error Mitigation}

Since we are in the NISQ era, rather then eliminating all the noise, the goal of different techniques is to rather mitigate them, therefore the field of quantum error mitigation was born. Multiple techniques have been evolved and are now used in order to extract useful information in real-world applications \cite{Endo_2018}, such as the ones related to quantum chemistry \cite{K_hn_2019, Lepp_kangas_2023, McCaskey_Parks_Jakowski_Moore_Morris_Humble_Pooser_2019}. Namely, some relevant techniques are zero noise extrapolation, probabilistic error cancellation, and error measurement mitigation \cite{QEM_Cai_2023}. Here, we will discuss Readout error mitigation and zero noise extrapolation. Dynamical Decoupling has also been tested, but the results were inconclusive as to whether they were more accurate.

\subsubsection{Readout Error Mitigation}
Readout Error Mitigation is an umbrella term used for multiple approaches to mitigating errors, usually with the help of a confusion matrix by applying its inverse to the outcome probability distribution to extract the noiseless distribution \cite{electronics11192983, Yang_2022}. This confusion matrix is a matrix that encodes the probabilities of measuring a prepared state $\ket{u}$ as another state, $\ket{v}$. One can consider whether the qubits' results are correlated between qubits or not. When correlated results are considered, for $n$ qubits, the matrix will be $2^n \times 2^n$. If the noise is, however, considered, local, we will have n matrices $2\times2$, one for each qubit.  
\par Ideally, regarding the type of error that is considered, this matrices will have one on the main diagonal, and 0 everywhere else, meaning that, the state prepared is always the same as the state measured. However, in a noisy environment this is not the case, therefore this matrix will encode the effects of noise on the states of the system. One simple way to construct this matrix is to prepare all the states $\ket{u}$ and then use the measurement distribution to construct its probabilities. This technique is, however, circuit independent. This means the circuit that will be run is not taken into consideration when performing these measurements. 
\par To extract the noiseless probability distribution, usually the inverse of the confusion matrix is calculated (the Moore-Penrose pseudoinverse) and then applied to the extracted probability distribution.
\subsubsection{Zero Noise Extrapolation}

Firstly introduced in 2017, Zero Noise Extrapolation is an error mitigation technique based on trying to extrapolate the noiseless result of a quantum circuit by analysing the circuit in different noise levels \cite{Giurgica_Tiron_2020, ZNE_2017, EfficientSimulatorZNE}. More technically, we need to estimate the expectation value of some observable with respect to an evolved state that is subject to noise. As the aforementioned article explains, they used Richardson extrapolation in order to extrapolate the zero-noise limit in short depth quantum circuits. 

With this being said, let $\lambda$ be the factor by which we increase the noise level. Following the notation from \cite{mitiq_LaRose_2022}, we let $\tau$ quantify the noise level in the circuit and let $\tau^\prime = \lambda\tau$ the scaled noise level. We can, therefore, see that for $\lambda = 1$, the input circuit remains unchanged. We let $\rho(\tau)$ be the density matrix representing the state prepared by the noise scaled quantum circuit. The expectation value of an observable A can therefore be represented as:

\begin{equation}
   \langle E(\lambda) \rangle = Tr[\rho(\lambda \tau) A]
\end{equation}

Therefore, if we measure for different values of $ \lambda \geq 1 $ we can try to extrapolate the zero-noise limit. Different functions can be used to extrapolate the result, the ones provided by \cite{mitiq_LaRose_2022} and that were used in the experiments are Richardson Extrapolation based and Polynomial extrapolation. Noise amplification, that means modifying the initial circuit for increased levels of $\lambda$ can be done through gate folding on a global or local scale \cite{Giurgica_Tiron_2020, majumdar2023best}. We used local folding for our circuits. Pulse level gate implementations can also be used in order to control the noise introduced \cite{pulse_streching1, pulse_streching2}.


\subsection{Multiprogramming - An Efficient Use of the Quantum Chip}
\par
\begin{figure}[h]
    \centering
    \includegraphics[scale=0.25]{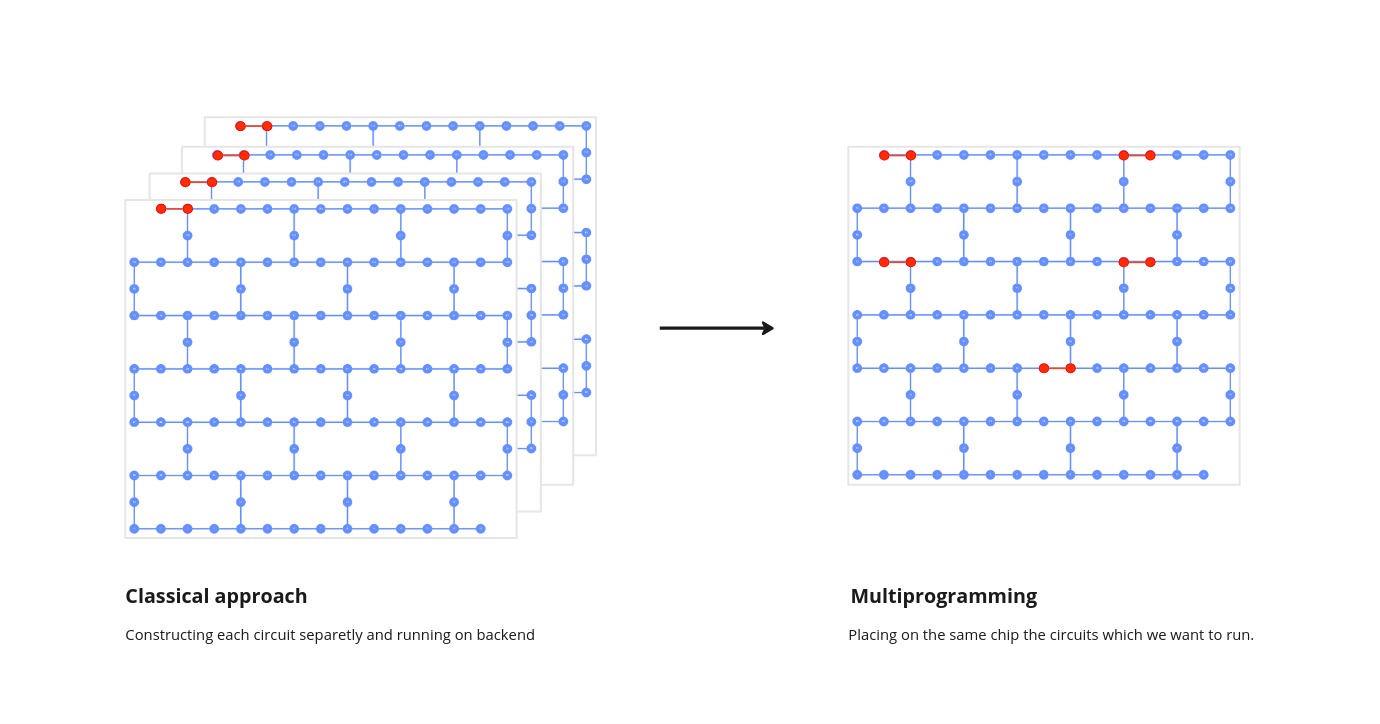}
    \caption{Multiprogamming allows for multiple circuits to run on the same chip}
    \label{fig:mp-scheme}
\end{figure}

 Whenever we want to run a certain experiment, one can take into consideration the efficient use of the quantum chip. Multiprogramming \cite{mp_Case_For_mp_2019, mp_Murali_2020, mp_Niu_2023, mp_Ohkura_2022} is a technique used for tackling the problem of hardware under-utilization in the NISQ era. Due to the circuits that are able to not be as affected by noise on the backend being restricted to smaller sizes and the hardware currently available, one would end up using only a small percentage of the qubits available in one run. For example, if we use the 2 qubit experiment, we want to run, and the chip of 128 qubits available through IBM, sending the experiments sequentially would result in the utilization of hardware of just $ 1.25\% $. Therefore, in our experiments, we chose to use multiprogramming for this reason. The main idea of this technique is highlighted in Figure \ref{fig:mp-scheme}.

\par Despite the advantages, there are also some disadvantages to multiprogramming, one of which is \textit{interference}. These interferences may occur due to the crosstalk introduced by additional operations and qubit measurement operations. Different layout strategies have been tested in order to limit this crosstalk. For example, there is evidence that in practice, for concurrent runs, crosstalk does not affect shorter circuits and for most runs, a physical buffer of one qubit between circuits is enough to reduce the effects of interfernece drastically if the circuit depth doesn't increase above 30 CX \cite{mp_Ohkura_2022}. Layout strategies have been developed to therefore increase the usage of a chip \cite{mp_Niu_2023}, along with scheduling algorithms that follow the same aim of reducing crosstalk \cite{mp_Case_For_mp_2019, mp_Murali_2020}.
For count-experiments, extracting the result for each circuit is straightforward. For statevector simulations, one can trace out the qubits that are not needed with the use of a partial trace over the density matrix of the system. If Q is the register of interest and A are the qubits for the other circuits, the density matrix of the system Q is extracted as in Equation \ref{eq:partial-trace}. 
\begin{equation} \label{eq:partial-trace}
    \rho_{Q} \equiv Tr_{A} [\rho_{QA}]
\end{equation}

\section{End to End Run}
\begin{figure}[h]
    \centering
    \includegraphics[scale=0.20]{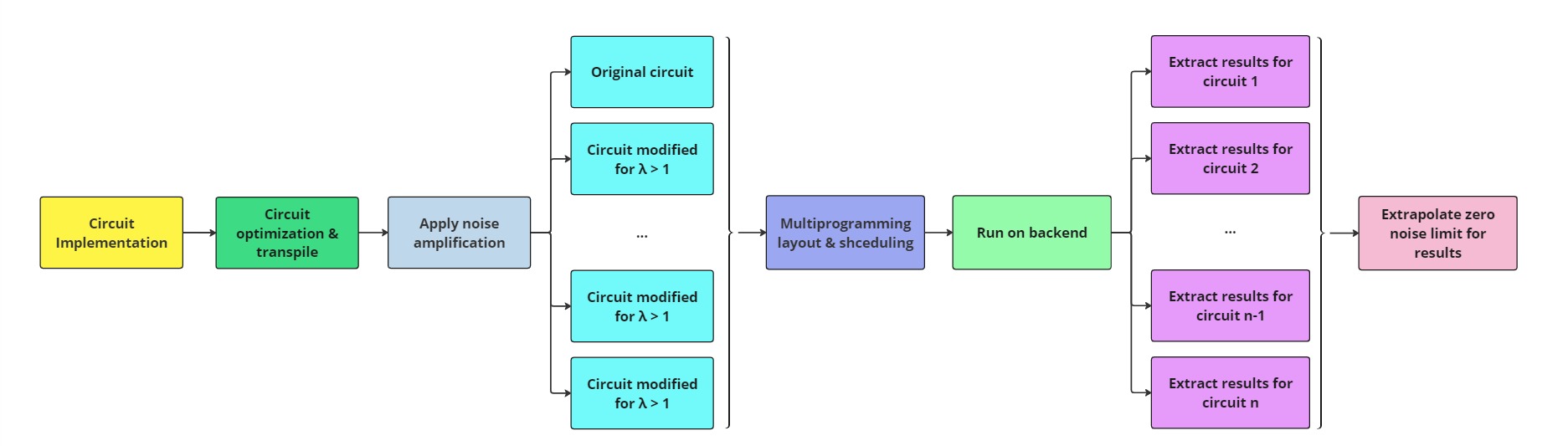}
    \caption{Workflow of the circuit}
    \label{fig:circuit-workflow}
\end{figure}

This experiment aims to \textit{simulate quantum tunneling} on a 2-qubit system. The barriers are placed on the highest order qubit, and the desired result is to see quantum tunneling between the two wells created. Moreover, we aim to \textit{optimize the circuit for running on hardware} and use error mitigation techniques (\textit{ZNE} in this case) to extract the best result possible of tunneling through the barrier. The \textit{multiprogramming} paradigm was employed to optimize hardware chip utilization. The steps of this experiment are illustrated in Figure \ref{fig:circuit-workflow}.

\subsubsection{Methodology}
For implementation, we used qiskit along with the tket compiler. For error mitigation techniques, we used mitiq \cite{mitiq_LaRose_2022}. For this experiment, the space was discretized using 2 qubits. We chose 7 timesteps and $\Delta t$ of 0.1. As hardware, the 128 qubit ibm\_osaka was used.

\subsubsection{Experimental Process}

\begin{figure}[H]
    \centering
    \includegraphics[scale=0.25]{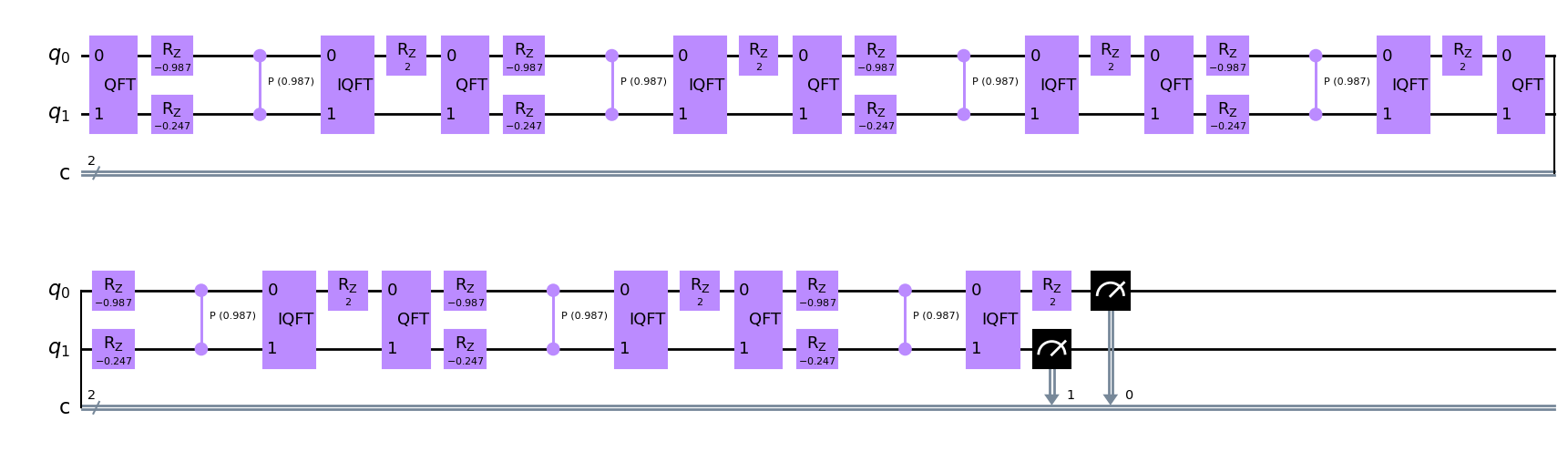}
    \caption{The circuit diagram}
    \label{fig:exp-end-to-end-circuit-diagram}
\end{figure}

Figure \ref{fig:exp-end-to-end-circuit-diagram} presents the abstract circuit diagram. Here the Quantum Fourier Transform operations were grouped in the gate QFT, the same for the inverse Fourier Transform. The single $Rz$ operator in between barriers represents the potential operator, and it implements the potential barrier. Mathematically, this operator is represented by (presented in section \ref{Potential-energy-operator}): 

$$
\hat{V} =  e^{-i v \sigma_z^{0} \Delta t} = I \otimes  e^{-i v \sigma_z \Delta t}
$$

Here, since we use qiskit for implementation we used qiskit ordering. This means that qubit 0 is the least-important qubit (uses little-endian convention). Therefore, this implements a potential barrier at $\ket{01}$ and $\ket{11}$.

\begin{figure}[H]
    \centering
    \begin{subfigure}[b]{0.45\textwidth}
        \centering
        \includegraphics[scale=0.20]{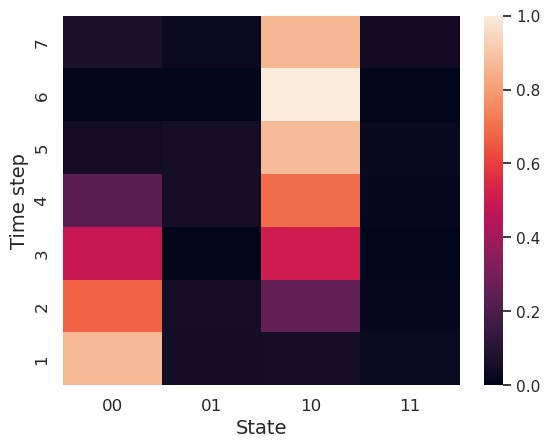}
        \caption{Plot of the 7 timesteps}
        \label{subfig:timesteps}
    \end{subfigure}
    \begin{subfigure}[b]{0.45\textwidth}
        \centering
        \includegraphics[scale=0.20]{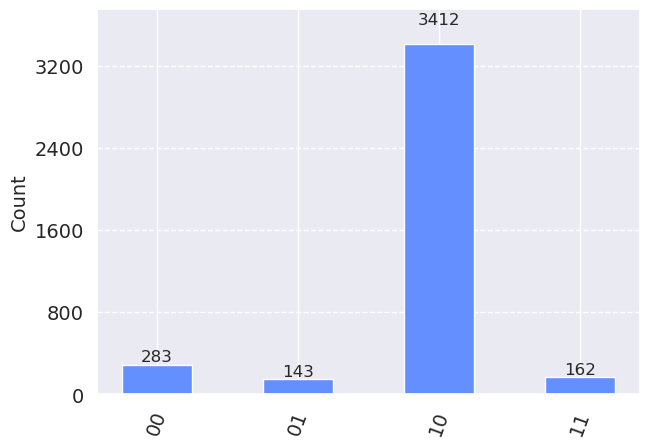}
        \caption{Ideal distribution for 4000 shots. Quantum tunneling is visible from well $\ket{00}$ to $\ket{10}$}
        \label{subfig:ideal_distribution}
    \end{subfigure}
    \caption{Experimental results for end-to-end process}
    \label{fig:exp-end-to-end-time-diagram}
\end{figure}

We can see the implemented evolution of the system in Figure \ref{fig:exp-end-to-end-time-diagram}. Tunneling is visible from well $\ket{00}$ to $\ket{10}$ (the particle starts at $\ket{00}$). This diagram shows the expected ideal output - it was calculated based on the density operators obtained by the statevectors taken at each timestep (based on ideal calculations, not noisy simulation or count distribution). Our wavefunction can now be written as $ \ket{\psi} = \sum_{i=0}^3 \alpha_i \ket{k_i} $, where $k_i$ is the statevector representing the binary representation of index $i$ and therefore the probability of being at a given $k_i$ is $\alpha_i ^2$

    
After the implementation of the original circuit, we needed to apply the noise amplification technique. In this experiment, we opted for \textit{local folding} and after conducting experiments with various factors, we determined scaling factors of $[1.0, 1.5, 2.0, 2.5, 3.0]$ to be most effective.
\par As for the layout technique for multiprogramming, we chose to leave a buffer of at least 1 physical qubit, following empirical evidence of other studies \cite{mp_Ohkura_2022}. We will also mention that the circuit has been compiled and prepared for backend run with the use of \textit{qiskit} as well as \textit{tket} compiler.

\subsubsection{Analysis of Experimental Results}
\par
\begin{figure}[H]
    \centering
    \begin{subfigure}[b]{\textwidth}
        \centering
        \includegraphics[scale=0.20]{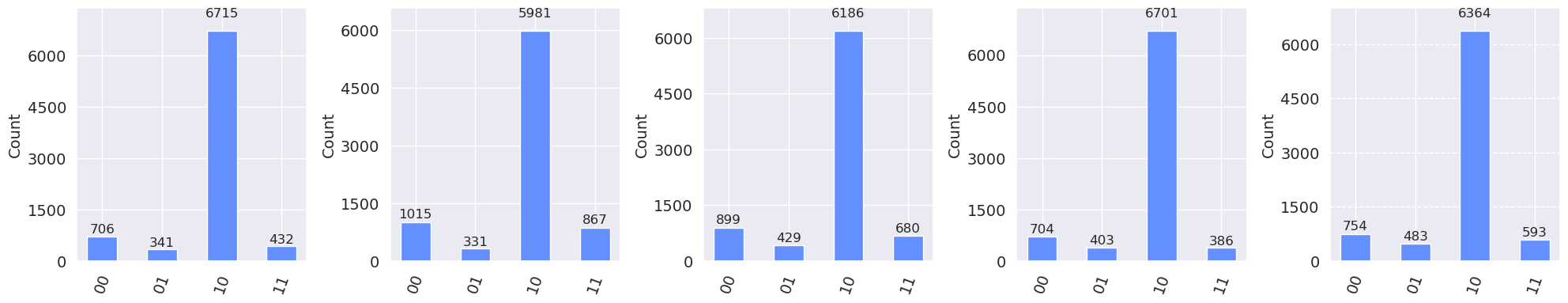}
        \caption{Results from backend run}
        \label{subfig:backend_results}
    \end{subfigure}
    \begin{subfigure}[b]{0.30\textwidth}
        \centering
        \includegraphics[scale=0.15]{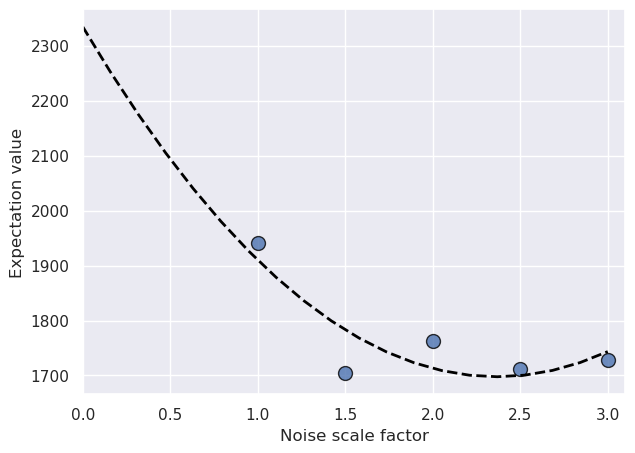}
        \caption{Inference on counts from noisy simulators}
        \label{subfig:noisy_simulations}
    \end{subfigure}
    \begin{subfigure}[b]{0.30\textwidth}
        \centering
        \includegraphics[scale=0.15]{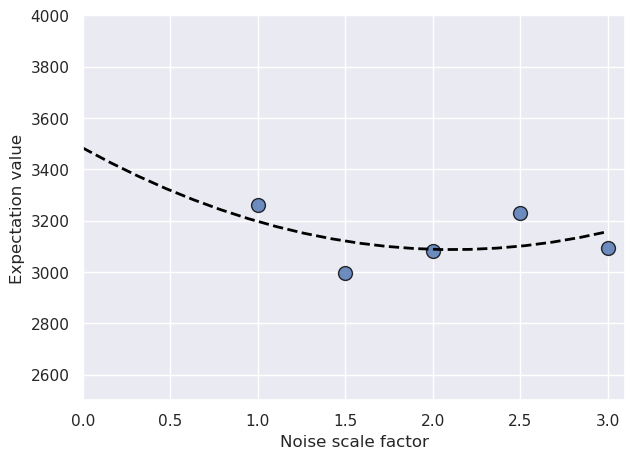}
        \caption{Inference on noisy results from backend}
        \label{subfig:noisy_backend_results}
    \end{subfigure}
    \caption{Experimental Results for End-to-End Process with Noise}
    \label{fig:exp-end-to-end-results}
\end{figure}

 Following the run on ibm\_osaka, the results for each noise amplified circuit are extracted and then used for inference of the zero noise limit. Results on noisy simulators are presented in Figure \ref{fig:exp-end-to-end-results}, along with the results from the circuits. Moreover, the results are concisely presented in Table \ref{tab:results-em}

    
\begin{table}[htbp]
    \centering
    \caption{Results on Noisy simulator and on Real Backend}
    \label{tab:results-em}
    
    \begin{center}
    \begin{tabular}{||c c c c c c||}  
         \hline
         Type & $T$ & $T_{run}$ & $E_1$ & $T_{run}^{em}$ & $E_2 $\\ [0.5ex] 
         \hline\hline
         simulation & 0.864 & 0.64 & 0.224 & 0.798 & 0.066  \\ 
         real run & 0.864 & 0.802 & 0.062 & 0.870 & 0.006  \\
         \hline
    \end{tabular}
    \end{center}
    
\end{table}
Where the following notations have been used:
\begin{conditions}
  T & the ideal probability of transmission; this entails finding the particle at location $\ket{10}$  \\
  T_{run} & The result for the run (either on a noisy simulator or on the backend) \\
  E_1 & unmitigated error, this is $T-T_{run}$ \\
  T_{run}^{em} & the transmission probability resulted from applying the error mitigation technique \\
  E_2 & The final error resulting after applying error mitigation, this is $|T-T_{run}^{em}|$
\end{conditions}
It is, therefore, evident that error mitigation had a crucial improvement over the simple, unmitigated run since we can observe that the inferred expectation value of our observable is only $0.006$ off. We can see that error mitigation played a crucial role in both simulated and real environments. In the simulated environments, error was reduced from $0.224$ to $0.066$, only a quarter of the original error. On the real run, the error was reduced from $0.062$ to $0.006$, observing yet again a substantial decrease. One can also appreciate the accuracy of the transmission probability achieved.   Moreover, using the multiprogramming paradigm, the chip was also efficiently used, since instead of using only 2 qubits out of the 128 qubits, five times in a row, all experiments were run on the same chip. 

\section{Extended Experimental Examination}
The setup outlined in this article demonstrates versatility, and we believe that conducting additional experiments will further enhance understanding regarding result interpretation and the various components of the workflow outlined. Therefore, we conducted 2 additional experiments, which will be presented in this section. The first one is a 2 qubit simulation using Readout Error Mitigation and the second one is a 6 qubit simulation in a noise-free environment that employs the Hadamard Test in order to get a better understanding of the wavefunction during quantum tunneling. 
\par For each experiment presented a timeline of probability will be firstly provided. A statevector simulator is used in order to get the probabilities at each timestep accurately. After the statevector from each timestep has been extracted, the probabilities are also extracted. 

\subsection{2 qubit simulation using REM}

\subsubsection{The objective(s) of the experiment}

The objective of the first experiment is to simulate quantum tunneling on a 2-qubit system. The barriers are placed on the highest order qubit, and the desired result is to see quantum tunneling between the two wells created. Moreover, we aim to optimize the circuit for running on hardware and compare the different optimization procedures for this experiment. Concisely, this experiment follows:
\begin{itemize}
    \item simulates quantum tunneling on a two-qubit system
    \item observe tunneling between the two wells
    \item optimize and run on hardware
    \item analyze the results
\end{itemize}

\subsubsection{Theoretical foundation: relevant theoretical concepts for the experiment}

For this example, implementation of the QFT, the potential and kinetic energy operators were needed. We used the implementation of the momentum operator without the use of an ancillary qubit since we didn't want to increase the depth of the circuit with swap gates required to mimic 3 qubit all-to-all connectivity.  

\subsubsection{Experiment Setup: Hardware, Software, and Key Parameters}

As hardware, we used the ibm\_belem, ibm\_nairobi and ibm\_osaka quantum computers, which have 5, 7 and 127 qubits, respectively. As of the 29th of November, the first two systems are retired, but we believe the results are still relevant due to the size of the experiment not being large. For the implementation, we used qiskit as the main SDK for implementing the operators discussed, along with the \textit{tket} compiler for further optimization and mapomatic for a more detailed result of the layout and mapping scores of the circuit.

\subsubsection{Discussion and Implications of Experimental Results}

\begin{figure}[H]
    \centering
    \includegraphics[scale=0.25]{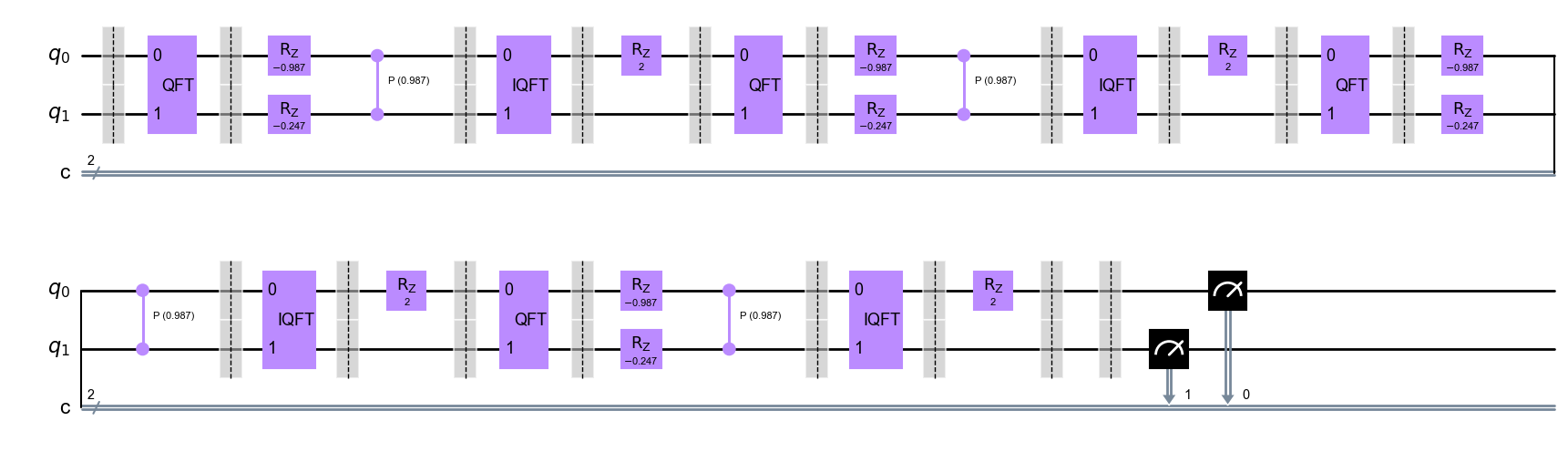}
    \caption{The circuit diagram}
    \label{fig:exp1-circuit-diagram}
\end{figure}

Firstly, we implemented the circuit with the help of qiskit. Figure \ref{fig:exp1-circuit-diagram} presents the abstract circuit diagram. Here the Quantum Fourier Transform operations were grouped in the gate QFT, the same for the inverse Fourier Transform. The single $Rz$ operator in between barriers represents the potential operator, and it implements the potential barrier. Mathematically, this operator is represented by (presented in section \ref{Potential-energy-operator}): 

$$
\hat{V} =  I \otimes e^{-i v \sigma_z^{0} \Delta t}
$$

Here, since we use qiskit for implementation we used qiskit ordering. This means that qubit 0 is the least-important qubit (uses little-endian convention). Therefore, this implements a potential barrier at $\ket{01}$ and $\ket{11}$.

\begin{figure}[H]
    \centering
    \includegraphics[scale=0.25]{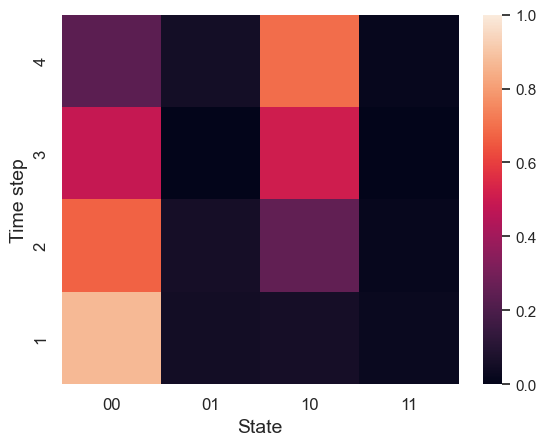}
    \caption{Plot of the 4 timesteps. Quantum tunneling is visible from well $\ket{00}$ to $\ket{10}$}
    \label{fig:exp1-time-diagram}
\end{figure}

The evolution of the implemented system is illustrated in Figure \ref{fig:exp1-time-diagram}. Tunneling is visible from well $\ket{00}$ to $\ket{10}$ (the particle starts at $\ket{00}$). This diagram shows the expected ideal output - it was calculated based on the density operators obtained by the statevectors taken at each timestep (based on ideal calculations, not noisy simulation or count distribution). Our wavefunction can now be written as $ \ket{\psi} = \sum_{i=0}^3 \alpha_i \ket{k_i} $, where $k_i$ is the statevector representing the binary representation of index $i$ and therefore the probability of being at a given $k_i$ is $\alpha_i ^2$.

\subsubsection{Analysis of Experimental Results}

\begin{figure}[H]
    \centering
    \begin{subfigure}[b]{0.45\textwidth}
        \centering
        \includegraphics[scale=0.15]{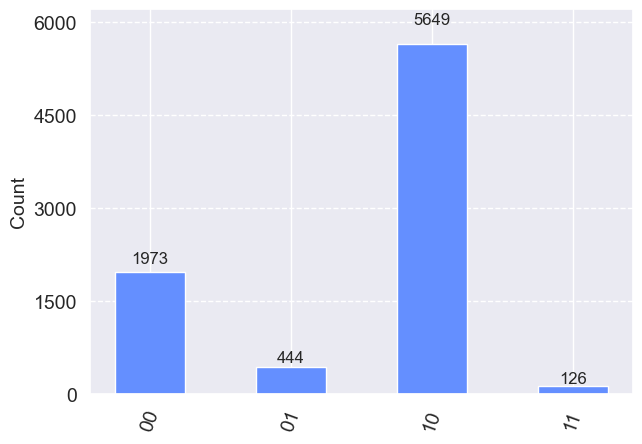}
        \caption{Count distribution}
        \label{subfig:count_distribution}
    \end{subfigure}
    \begin{subfigure}[b]{0.45\textwidth}
        \centering
        \includegraphics[scale=0.15]{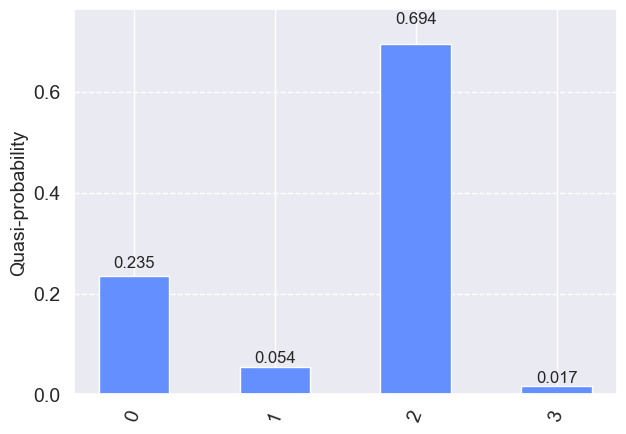}
        \caption{Sampler output}
        \label{subfig:sampler_output}
    \end{subfigure}
    \caption{Experimental results for the 2q experiment}
    \label{fig:exp1-sampler}
\end{figure}

Count distribution and probabilities from running on the qasm\_simulator and Sampler primitive are presented in Figure \ref{fig:exp1-sampler}. Measurements were made only after all steps had been executed, as seen in the circuit diagram in Figure \ref{fig:exp1-circuit-diagram}, so represented below are the probabilities of the last step. Therefore, we see that the particle has tunneled from well $\ket{00}$ to $\ket{10}$ by the probability given for state $\ket{10}$.

Firstly, we tried running our circuit on the qiskit backend without further intervention in the transpilation process. The hardware used in this experiment was the 5-qubit quantum computer named ibm\_belem (which has been retired) and the 127 quantum computer ibm\_osaka. The results of this run are presented in Figure \ref{fig:exp1-backend-no-em}.

\begin{figure}[H]
    \centering
    \includegraphics[scale=0.15]{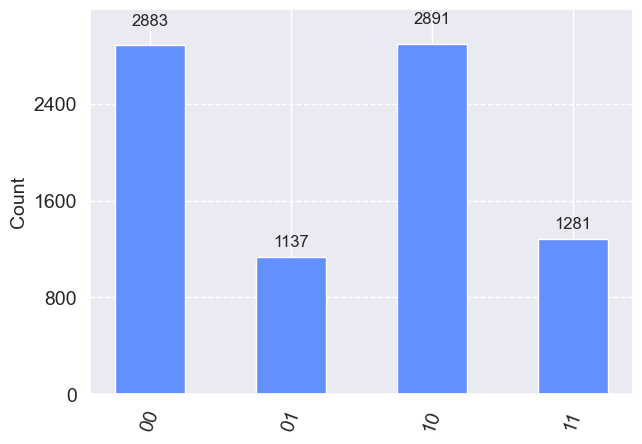}
    \caption{Backend run}
    \label{fig:exp1-backend-no-em}
\end{figure}

We can now notice the difference between simulations and running on hardware with the use of Figures \ref{fig:exp1-sampler} and \ref{fig:exp1-backend-no-em}. Running the circuit on quantum computers has the disadvantage of being affected by error, since we are now in the NISQ era of quantum computing. The error can be seen in the results, for example, as erroneous counts on states $\ket{01}$ and $\ket{11}$. Errors can have multiple factors, ranging from external noise affecting the states of the qubit to the routing and layout chosen by the transpiler not being the optimal solution for the given circuit. Because of this, we can try to optimise this circuit in order to make it more prone to errors. As presented in section \ref{Transpilation-passes}, we can further optimize the passes presented. For this experiment, we chose to use the \textit{tket} compiler as an add-on to the \textit{qiskit} optimization passes in order to optimize the circuit further. Moreover, for the layouting technique, we used the \textit{mapomatic} library and the routing from qiskit. By applying this optimization, the following result is obtained.


\begin{figure}[H]
    \centering
    \begin{subfigure}[b]{0.45\textwidth}
        \centering
        \includegraphics[scale=0.2]{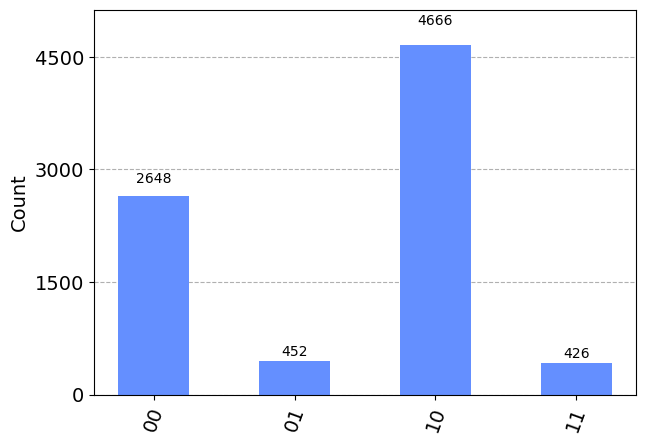}
        \caption{Backend run on Nairobi}
        \label{subfig:nairobi_backend}
    \end{subfigure}
    \begin{subfigure}[b]{0.45\textwidth}
        \centering
        \includegraphics[scale=0.2]{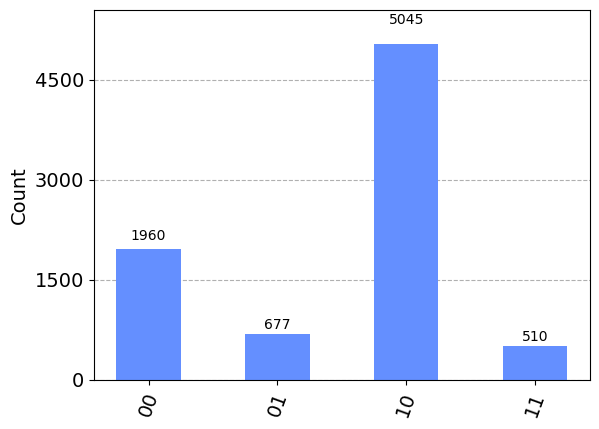}
        \caption{Backend run on Osaka}
        \label{subfig:osaka_backend}
    \end{subfigure}
    \caption{Experimental Results for 2 qubit simulation without error mitigation}
    \label{fig:exp1-backend-no-em}
\end{figure}

Applying the same optimization for another backend, for example, for the 127-qubit Osaka quantum computer, the result in Figure \ref{subfig:osaka_backend}

    

\subsubsection{Readout Error Mitigation}

\begin{figure}[H]
    \centering
    \begin{subfigure}[b]{0.45\textwidth}
        \centering
        \includegraphics[scale=0.2]{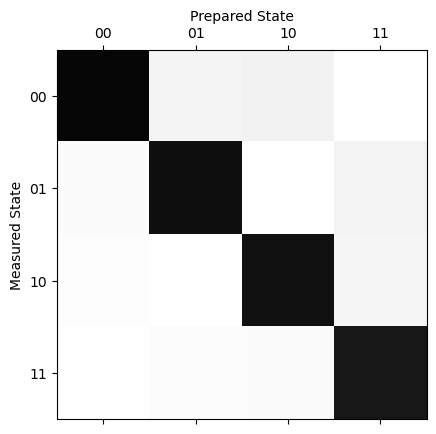}
        \caption{Confusion matrix}
        \label{subfig:confusion_matrix}
    \end{subfigure}
    \hspace{5mm}
    \begin{subfigure}[b]{0.45\textwidth}
        \centering
        \includegraphics[scale=0.2]{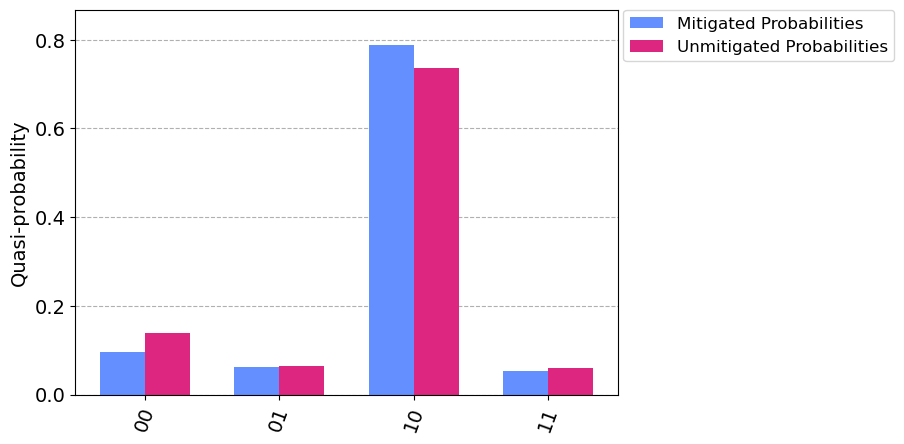}
        \caption{Probabilities compared}
        \label{subfig:probabilities_compared}
    \end{subfigure}
    \caption{Experimental Results for 2 qubit simulation with REM}
    \label{fig:exp1-rem}
\end{figure}

An error mitigation technique that can be used in practice is Readout Error Mitigation. It is based on constructing an assignment matrix similar to a confusion matrix, meaning it is constructed in order to show how often a result is measured wrongly when, in reality, the true outcome should be different. Local and Correlated readout error mitigators can be used, but here we chose to use Correlated readout error mitigators. Both the confusion matrix and the comparison are presented in Figure \ref{fig:exp1-rem}. The confusion matrix observed in Figure \ref{subfig:confusion_matrix} has been constructed with the use of simple circuits (by this, we mean minimal in-depth, by using only additional X gates when needed) preparing each possible 2-qubit state. We can see in Figure \ref{subfig:probabilities_compared} the results of the mitigated run. While the difference is not substantial, improvements can still be observed, for example, in positions where barriers should be.

\subsection{6 qubit Quantum Tunneling Simulation using the Hadamard Test} \label{sec:hadamard-test-experiment}

\subsubsection{The objective(s) of the experiment}

The objective of the experiment is to simulate quantum tunneling in a 6-qubit system. Here, the Hadamard test was used in order to "extract" (estimate) the real and imaginary part of the wavefunction rather than the probability of said wavefunction. Concisely, the objectives of the experiment were:
\begin{itemize}
    \item implements the simulation circuit in a bigger 6-qubit system
    \item extract probability counts, observe tunneling
    \item extract real and imaginary part counts, observe how they evolve
\end{itemize}

\subsubsection{Theoretical foundation: relevant theoretical concepts for the experiment}

Apart from the QFT, potential, and kinetic energy operators, the Hadamard test is also needed for this experiment. Therefore, in order to simulate in one run and interpret the result, familiarity with how the Hadamard test works is needed.

\textbf{Hadamard test} 

\begin{figure}
    \centering
    \includegraphics[scale=0.45]{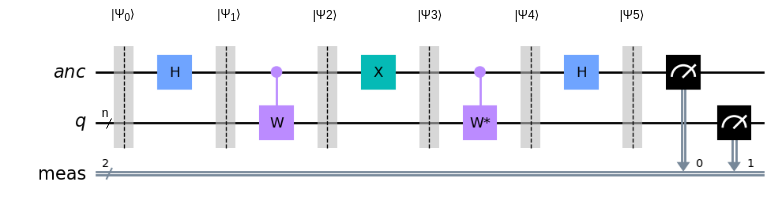}    
    \caption{Hadamard test implementation}
    \label{fig:fig_hadamrd_general}
\end{figure}

The Hadamard test is used in case we want to extract the distribution of  $Re\{\psi(x, t)\}^2$ and $Im\{\psi(x, t)\}^2$ separately. This is done with the help of controlled operations. Suppose we have a state of superposition:
$$
    \ket{\psi} = \sum_i \alpha_i \ket{k_i}
$$
where $\alpha_i$ is, of course, a complex number and $\ket{k}$ represent the base states. Let $W$ be the operator that takes the state from $\ket{0}^{\otimes n}$ to $\ket{\psi}$, that being $\ket{\psi} = W \ket{0}^{\otimes n}$. In order to extract the distribution of the real and imaginary parts, we will also use the complex conjugate of the operator $W$, $W^*$. An auxiliary qubit is also added. The Hadamard test is therefore represented:
\begin{enumerate}
    \item Initialize the system in the state $\ket{\psi_0}=\ket{0}\ket{0}^{\otimes n}$
    \item Apply Hadamard to ancillary qubit: $\ket{\psi_1}=\ket{+}\ket{0}^{\otimes n}$
    \item Apply Controlled $W$ operator: $\ket{\psi_2}=\frac{1}{\sqrt{2}}(\ket{0}W\ket{0}^{\otimes n} + \ket{1}\ket{0}^{\otimes n}) = \frac{1}{\sqrt{2}}(\ket{0}\ket{\psi} + \ket{1}\ket{0}^{\otimes n})$
    \item Apply Not gate on ancillary qubit $\ket{\psi_3}=\frac{1}{\sqrt{2}}(\ket{1}\ket{\psi} + \ket{0}\ket{0}^{\otimes n})$
    \item Apply Controlled $W^*$ operator: $\ket{\psi_4}=\frac{1}{\sqrt{2}}(\ket{1}\ket{\psi} + \ket{0}W^*\ket{0}^{\otimes n}) = \frac{1}{\sqrt{2}}(\ket{1}\ket{\psi} + \ket{0}\ket{\psi^*})$, where $W^*\ket{0}^{\otimes n} = \ket{\psi^*}$
    \item Apply Hadamrd operator again $\ket{\psi_5} = \frac{1}{2}(\ket{0}\ket{\psi} - \ket{1}\ket{\psi} + \ket{0}\ket{\psi^*} + \ket{1}\ket{\psi^*}) = \frac{1}{2}(\ket{0}(\ket{\psi} + \ket{\psi^*}) + \ket{1}(\ket{\psi^*} - \ket{\psi}))$
    \item Measuring the qubits, it can be observed that if the ancilla qubit is measured as 0, then the result from the second register defines $Re\{\psi(x, t)\}^2$, while if the ancilla qubit is 1, the $Im\{\psi(x, t)\}^2$ distribution can be extracted.
\end{enumerate}
Schematically, the implementation is represented in Figure \ref{fig:fig_hadamrd_general}

\subsubsection{Experiment Setup: Hardware, Software, and Key Parameters}

\begin{figure}[H]
    \centering
    \includegraphics[scale=0.4]{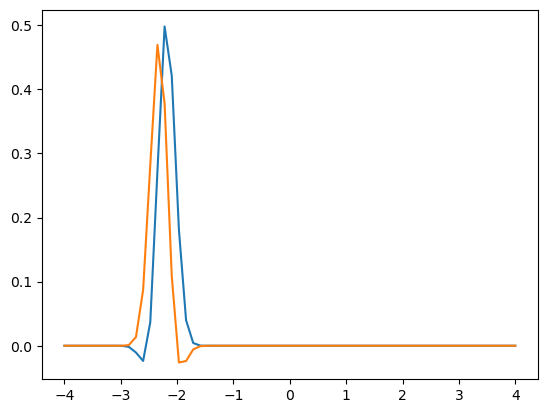}
    \caption{Initial wavepacket}
    \label{fig:exp3-initial-wave}
\end{figure}

Here we used qiskit for implementation. Moreover, we didn't need the tket compiler for hardware optimisation, since the depth of the circuit wouldn't allow for current free available quantum computers to run. Since the space here allows it, we will use as the initial wavefunction the Gaussian function. Since the depth of this circuit is large, we restrained to only simulating it on qasm\_simulator. The initial wavepacket is represented as:

$$
\ket{\psi} = \frac{1}{\sqrt{2\pi} 0.8} e^{-\frac{1}{2}(\frac{x+2.25}{0.032})^2} e^{-300ix}
$$
That is to say, the potential of the wave is 300. This number was chosen because we wanted a wave that traveled to the right. Mean, and standard deviations were chosen for the wavefunction to start before the barrier and not extend beyond it. The initial wavepacket is presented in Figure \ref{fig:exp3-initial-wave}.

\subsubsection{Discussion and Implications of Experimental Results}

\begin{figure}[H]
    \centering
    \includegraphics[scale=0.25]{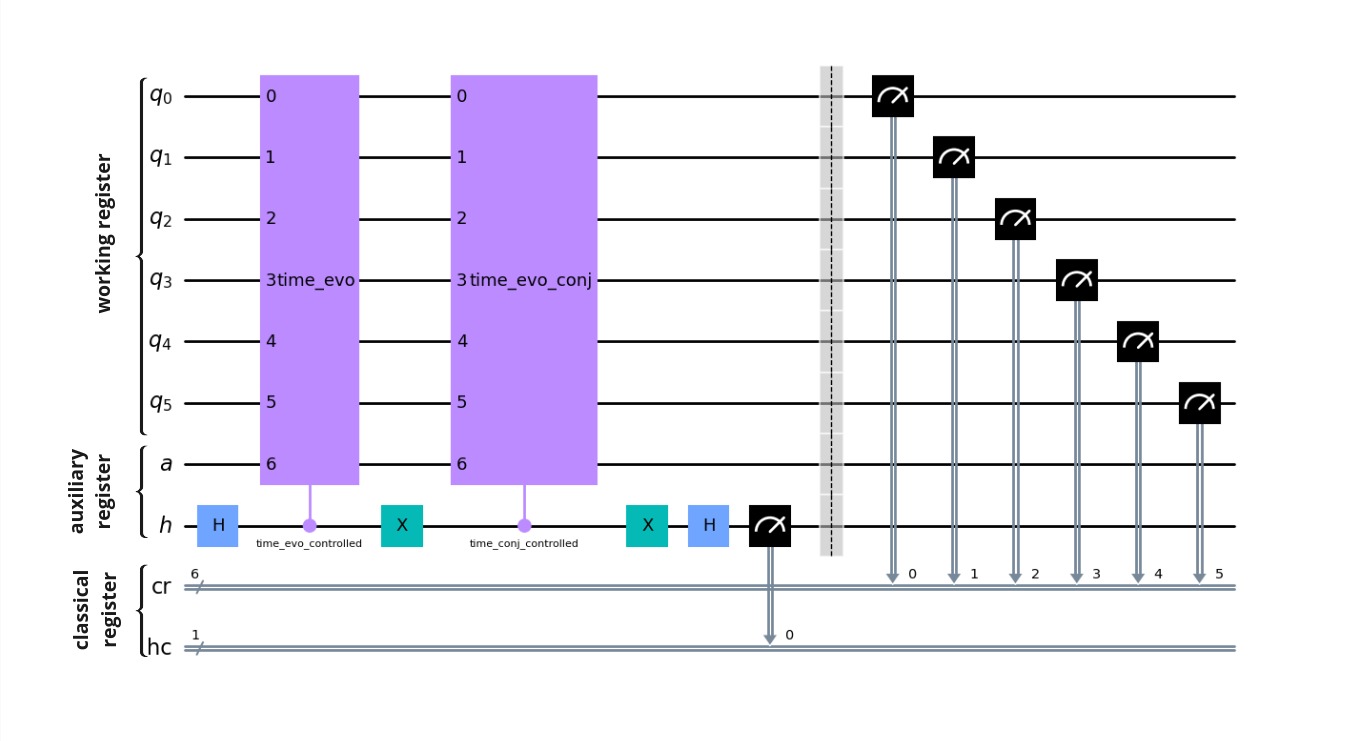} 
    \caption{Implementation of Hadamard test. The working register \textit{q} is of size 6. These are the qubits for which the operators were used. The auxiliary register contains one qubit \textit{a}, which is used for the implementation of the momentum operator, and a qubit \textit{h} needed for the Hadamard test - depending on this value, the operator or the conjugate operator was applied in order to extract the real/imaginary part. \textit{cr} are the classical bits that hold the measurements of the 6 qubits. \textit{hc} is of size 1. This holds the value of the auxiliary qubit \textit{h}. Depending on this value, we know that we measured the imaginary or real part of the wavefunction.}
    \label{fig:had-test-6-qubit}
\end{figure}

As presented in the Hadamard test section, the implementation of this type of function requires controlled operations and also an auxiliary qubit as the control qubit. The implementation of the state preparation, QFT, IQFT, potential and kinetic energy operators were grouped into the gate \textit{time\_evo}. We should mention that this includes all the timesteps from the troterrized form. The implementation of the momentum operator followed the schema with the auxiliary qubit, here noted as \textit{a}. The circuit is presented in Figure \ref{fig:had-test-6-qubit}. 

\subsubsection{Analysis of Experimental Results}

\begin{figure}[H]
    \centering
    \begin{subfigure}[b]{0.45\textwidth}
        \centering
        \includegraphics[scale=0.15]{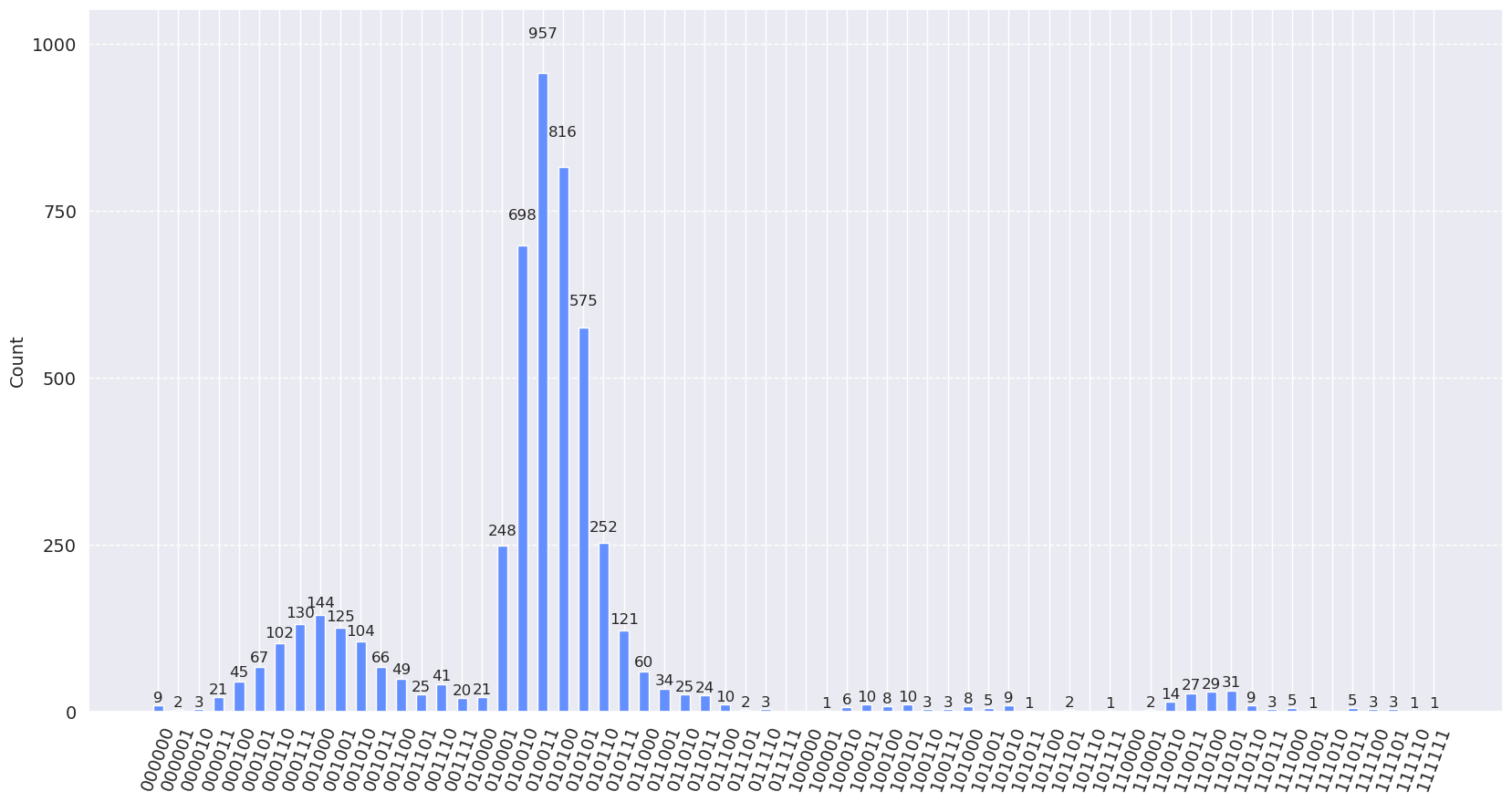}
        \caption{Counts probability distribution}
        \label{subfig:prob_distribution}
    \end{subfigure}
    \begin{subfigure}[b]{0.45\textwidth}
        \centering
        \includegraphics[scale=0.15]{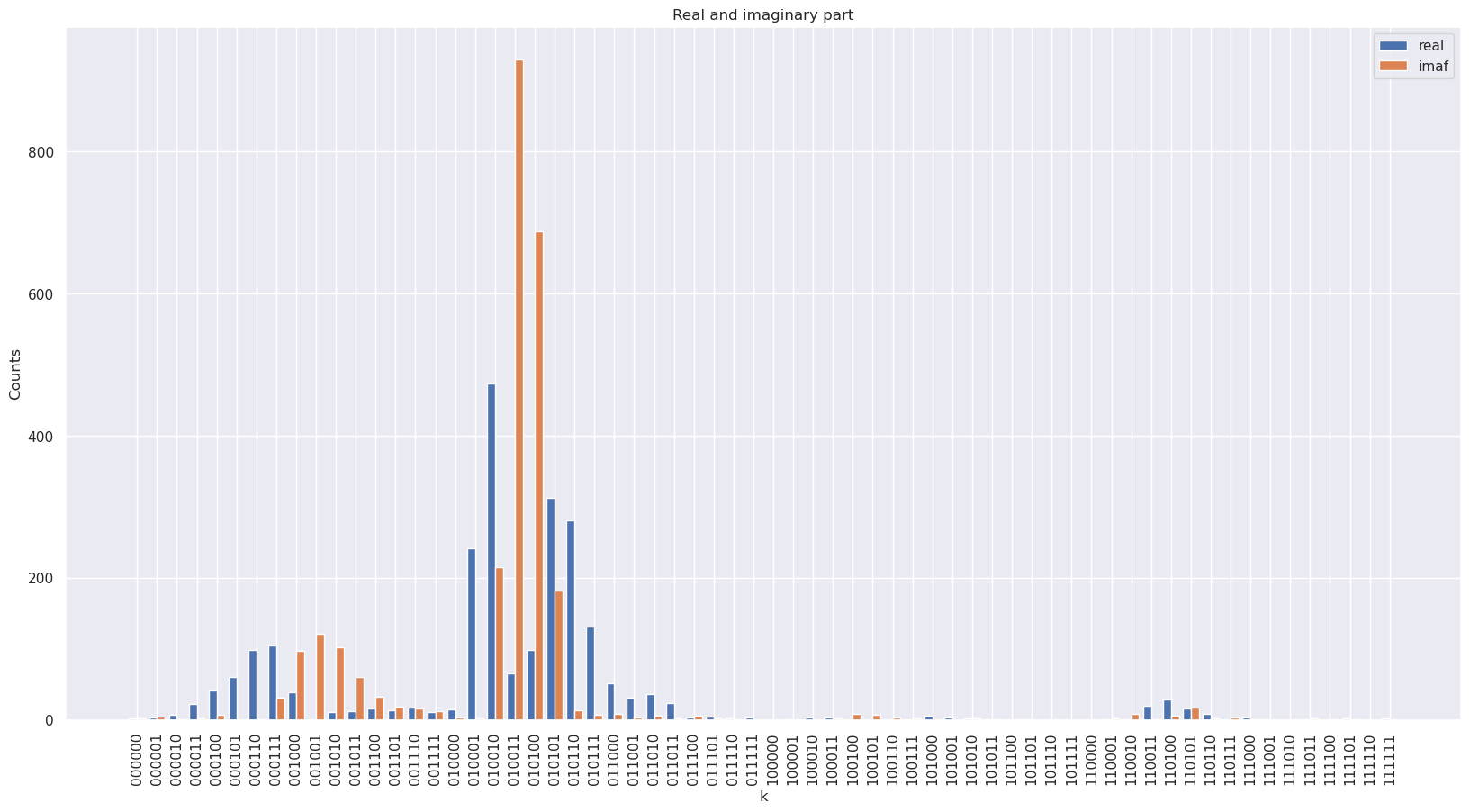}
        \caption{Real and imaginary count distributions}
        \label{subfig:real_imag_distribution}
    \end{subfigure}
    \caption{Experimental Results for Hadamard Test Experiment}
    \label{fig:had-test-prob-distr}
\end{figure}

The simulation was run on qasm\_simulator and the Figure \ref{fig:had-test-prob-distr} shows probability distribution and imaginary and real counts distribution of the evolved state. The potential barrier is placed at positions xxx11x (the implementation of this potential barrier is discussed in Section \ref{sec:four-qub-quant-sim}), and we can clearly see tunneling of the wavepacket after the potential barrier. Moreover, we can observe the real and imaginary distributions in Figure \ref{subfig:real_imag_distribution}. The transmission wave is clearly illustrated in this image, and through these larger simulations in terms of qubits used, which means a more fine discretization of space, one can try to approach real simulations more accurately. 

\section{Conclusions}
 In this article, we have outlined the primary procedures necessary for translating the quantum mechanical challenge of simulating the Schr\"{o}dinger equation into a quantum computing framework. That is, we began by examining the theoretical implications of simulating quantum tunneling and subsequently deliberated on diverse methodologies for operator implementation.  
\par Moreover, we focused on the changes required in order to efficiently and accurately run the abstract circuit on a real backend. We discussed what the transpilation procedure entails and why a call for hardware-aware design is still necessary in the NISQ era in the context of superconducting architectures. Multiple compilers have been employed in order to optimize the circuit for the chosen backend, and they proved capable of efficient optimization. 
Error mitigation has also been discussed, and ZNE and REM are used to improve the simulation's results. For today's chip's hardware underutilization problem, multiprogramming was harnessed to optimally run the circuit along with the noise-amplified circuits alongside the aforementioned error mitigation techniques. Hadamard tests for 6-qubit systems were also simulated to enrich the understanding of the topic. 
 \par Our research yields multifaceted implications. Through the successful simulation of quantum tunneling within a comprehensive workflow, we achieve precise results in a two-qubit simulation. We assert that this adaptable workflow offers significant value across diverse simulation contexts, thereby providing universal benefits to researchers.

\nocite{*}
\bibliographystyle{plain} 
\bibliography{refs} 

\begin{thebibliography}{10}

\bibitem{Abouelela2020}
Mohamed Abouelela.
\newblock Quantum simulation of the schrodinger equation using ibm’s quantum
  computers.
\newblock Bachelor's thesis is physics, 2020.

\bibitem{electronics11192983}
Martin Beisel, Johanna Barzen, Frank Leymann, Felix Truger, Benjamin Weder, and
  Vladimir Yussupov.
\newblock Configurable readout error mitigation in quantum workflows.
\newblock {\em Electronics}, 11(19), 2022.

\bibitem{Benenti_2008}
Giuliano Benenti and Giuliano Strini.
\newblock Quantum simulation of the single-particle schrödinger equation.
\newblock {\em American Journal of Physics}, 76(7):657--662, jul 2008.

\bibitem{QEM_Cai_2023}
Zhenyu Cai, Ryan Babbush, Simon~C. Benjamin, Suguru Endo, William~J. Huggins,
  Ying Li, Jarrod~R. McClean, and Thomas~E. O’Brien.
\newblock Quantum error mitigation.
\newblock {\em Reviews of Modern Physics}, 95(4), December 2023.

\bibitem{Cirac2012GoalsAO}
J.~Ignacio Cirac and Peter Zoller.
\newblock Goals and opportunities in quantum simulation.
\newblock {\em Nature Physics}, 8:264 -- 266, 2012.

\bibitem{Daley2022PracticalQA}
Andrew~J. Daley, Immanuel Bloch, C.~Kokail, Stuart Flannigan, N~Pearson,
  Matthias Troyer, and Peter Zoller.
\newblock Practical quantum advantage in quantum simulation.
\newblock {\em Nature}, 607:667 -- 676, 2022.

\bibitem{mp_Case_For_mp_2019}
Poulami Das, Swamit~S. Tannu, Prashant~J. Nair, and Moinuddin Qureshi.
\newblock A case for multi-programming quantum computers.
\newblock In {\em Proceedings of the 52nd Annual IEEE/ACM International
  Symposium on Microarchitecture}, MICRO '52, page 291–303, New York, NY,
  USA, 2019. Association for Computing Machinery.

\bibitem{Trotter_Suzuki_1}
Ish Dhand and Barry~C Sanders.
\newblock Stability of the trotter–suzuki decomposition.
\newblock {\em Journal of Physics A: Mathematical and Theoretical},
  47(26):265206, June 2014.

\bibitem{Endo_2018}
Suguru Endo, Simon~C. Benjamin, and Ying Li.
\newblock Practical quantum error mitigation for near-future applications.
\newblock {\em Physical Review X}, 8(3), July 2018.

\bibitem{Feynman_1982}
Richard~P. Feynman.
\newblock Simulating physics with computers.
\newblock {\em International Journal of Theoretical Physics},
  21(6–7):467–488, Jun 1982.

\bibitem{frança2022efficient}
Daniel~Stilck França, Liubov~A. Markovich, V.~V. Dobrovitski, Albert~H.
  Werner, and Johannes Borregaard.
\newblock Efficient and robust estimation of many-qubit hamiltonians, 2022.

\bibitem{pulse_streching2}
J.~W.~O. Garmon, R.~C. Pooser, and E.~F. Dumitrescu.
\newblock Benchmarking noise extrapolation with the openpulse control
  framework.
\newblock {\em Phys. Rev. A}, 101:042308, Apr 2020.

\bibitem{Giurgica_Tiron_2020}
Tudor Giurgica-Tiron, Yousef Hindy, Ryan LaRose, Andrea Mari, and William~J.
  Zeng.
\newblock Digital zero noise extrapolation for quantum error mitigation.
\newblock In {\em 2020 IEEE International Conference on Quantum Computing and
  Engineering (QCE)}. IEEE, October 2020.

\bibitem{Greenaway_2024}
Sean Greenaway, Adam Smith, Florian Mintert, and Daniel Malz.
\newblock Analogue quantum simulation with fixed-frequency transmon qubits.
\newblock {\em Quantum}, 8:1263, February 2024.

\bibitem{Griffiths_Schroeter_2018}
David~J. Griffiths and Darrell~F. Schroeter.
\newblock {\em Introduction to Quantum Mechanics}.
\newblock Cambridge University Press, 3 edition, 2018.

\bibitem{Trotter_Suzuki_2}
Naomichi Hatano and Masuo Suzuki.
\newblock {\em Finding Exponential Product Formulas of Higher Orders}, page
  37–68.
\newblock Springer Berlin Heidelberg, November 2005.

\bibitem{hegade2021experimental}
Narendra~N. Hegade, Nachiket~L. Kortikar, Bikramaditya Das, Bikash~K. Behera,
  and Prasanta~K. Panigrahi.
\newblock Experimental demonstration of quantum tunneling in ibm quantum
  computer, 2021.

\bibitem{pulse_streching1}
Ivan Henao, Jader Santos, and Raam Uzdin.
\newblock Adaptive quantum error mitigation using pulse-based inverse
  evolutions.
\newblock {\em npj Quantum Information}, 9, 11 2023.

\bibitem{Huang_2020}
He-Liang Huang, Dachao Wu, Daojin Fan, and Xiaobo Zhu.
\newblock Superconducting quantum computing: a review.
\newblock {\em Science China Information Sciences}, 63(8), July 2020.

\bibitem{quantum-sim-qho-2019}
Valay Jain, Bikash Behera, and Prasanta Panigrahi.
\newblock Quantum simulation of discretized harmonic oscillator on ibm quantum
  computer.
\newblock 07 2019.

\bibitem{kumar2019quantum}
Shubham Kumar, Rahul~Pratap Singh, Bikash~K. Behera, and Prasanta~K. Panigrahi.
\newblock Quantum simulation of negative hydrogen ion using variational quantum
  eigensolver on ibm quantum computer, 2019.

\bibitem{K_hn_2019}
Michael Kühn, Sebastian Zanker, Peter Deglmann, Michael Marthaler, and Horst
  Weiß.
\newblock Accuracy and resource estimations for quantum chemistry on a
  near-term quantum computer.
\newblock {\em Journal of Chemical Theory and Computation}, 15(9):4764–4780,
  August 2019.

\bibitem{mitiq_LaRose_2022}
Ryan LaRose, Andrea Mari, Sarah Kaiser, Peter~J. Karalekas, Andre~A. Alves,
  Piotr Czarnik, Mohamed El~Mandouh, Max~H. Gordon, Yousef Hindy, Aaron
  Robertson, Purva Thakre, Misty Wahl, Danny Samuel, Rahul Mistri, Maxime
  Tremblay, Nick Gardner, Nathaniel~T. Stemen, Nathan Shammah, and William~J.
  Zeng.
\newblock Mitiq: A software package for error mitigation on noisy quantum
  computers.
\newblock {\em Quantum}, 6:774, August 2022.

\bibitem{Lepp_kangas_2023}
Juha Leppäkangas, Nicolas Vogt, Keith~R. Fratus, Kirsten Bark, Jesse~A.
  Vaitkus, Pascal Stadler, Jan-Michael Reiner, Sebastian Zanker, and Michael
  Marthaler.
\newblock Quantum algorithm for solving open-system dynamics on quantum
  computers using noise.
\newblock {\em Physical Review A}, 108(6), December 2023.

\bibitem{EfficientSimulatorZNE}
Ying Li and Simon~C. Benjamin.
\newblock Efficient variational quantum simulator incorporating active error
  minimization.
\newblock {\em Phys. Rev. X}, 7:021050, Jun 2017.

\bibitem{liao2023machine}
Haoran Liao, Derek~S. Wang, Iskandar Sitdikov, Ciro Salcedo, Alireza Seif, and
  Zlatko~K. Minev.
\newblock Machine learning for practical quantum error mitigation, 2023.

\bibitem{Calibration_Liu}
Yiding Liu, Zedong Li, Alan Robertson, Xin Fu, and Shuaiwen~Leon Song.
\newblock Enabling efficient real-time calibration on cloud quantum machines.
\newblock {\em IEEE Transactions on Quantum Engineering}, 4:1--17, 2023.

\bibitem{Lloyd1996UniversalQS}
Seth Lloyd.
\newblock Universal quantum simulators.
\newblock {\em Science}, 273:1073 -- 1078, 1996.

\bibitem{QuantumDots_Divicenzo}
Daniel Loss and David~P. DiVincenzo.
\newblock Quantum computation with quantum dots.
\newblock {\em Phys. Rev. A}, 57:120--126, Jan 1998.

\bibitem{majumdar2023best}
Ritajit Majumdar, Pedro Rivero, Friederike Metz, Areeq Hasan, and Derek~S Wang.
\newblock Best practices for quantum error mitigation with digital zero-noise
  extrapolation, 2023.

\bibitem{McCaskey_Parks_Jakowski_Moore_Morris_Humble_Pooser_2019}
Alexander~J. McCaskey, Zachary~P. Parks, Jacek Jakowski, Shirley~V. Moore,
  Titus~D. Morris, Travis~S. Humble, and Raphael~C. Pooser.
\newblock Quantum chemistry as a benchmark for near-term quantum computers.
\newblock {\em npj Quantum Information}, 5(1), Nov 2019.

\bibitem{mp_Murali_2020}
Prakash Murali, David~C. Mckay, Margaret Martonosi, and Ali Javadi-Abhari.
\newblock Software mitigation of crosstalk on noisy intermediate-scale quantum
  computers.
\newblock In {\em Proceedings of the Twenty-Fifth International Conference on
  Architectural Support for Programming Languages and Operating Systems},
  ASPLOS ’20. ACM, March 2020.

\bibitem{QuantumComputationAndInformation}
Michael~A. Nielsen and Isaac~L. Chuang.
\newblock {\em Quantum Computation and Quantum Information: 10th Anniversary
  Edition}.
\newblock Cambridge University Press, 2011.

\bibitem{mp_Niu_2023}
Siyuan Niu and Aida Todri-Sanial.
\newblock Enabling multi-programming mechanism for quantum computing in the
  nisq era.
\newblock {\em Quantum}, 7:925, February 2023.

\bibitem{mp_Ohkura_2022}
Yasuhiro Ohkura, Takahiko Satoh, and Rodney Van~Meter.
\newblock Simultaneous execution of quantum circuits on current and near-future
  nisq systems.
\newblock {\em IEEE Transactions on Quantum Engineering}, 3:1–10, 2022.

\bibitem{Preskill_2018}
John Preskill.
\newblock Quantum computing in the nisq era and beyond.
\newblock {\em Quantum}, 2:79, August 2018.

\bibitem{Rodrigues2018}
Afonso Rodrigues.
\newblock {\em Validation of quantum simulations: Assessing efficiency and
  reliability in experimental implementations}.
\newblock PhD thesis, Universidade do Minho, 2018.

\bibitem{Shokri_2021}
Sina Shokri, Shahnoosh Rafibakhsh, Roghayeh Pooshgan, and Rita Faeghi.
\newblock Implementation of a quantum algorithm to estimate the energy of a
  particle in a finite square well potential on {IBM} quantum computer.
\newblock {\em The European Physical Journal Plus}, 136(7), jul 2021.

\bibitem{Sornborger_2012}
Andrew~T. Sornborger.
\newblock Quantum simulation of tunneling in small systems.
\newblock {\em Scientific Reports}, 2(1), aug 2012.

\bibitem{Tacchino_2019}
Francesco Tacchino, Alessandro Chiesa, Stefano Carretta, and Dario Gerace.
\newblock Quantum computers as universal quantum simulators: State-of-the-art
  and perspectives.
\newblock {\em Advanced Quantum Technologies}, 3(3), dec 2019.

\bibitem{ZNE_2017}
Kristan Temme, Sergey Bravyi, and Jay~M. Gambetta.
\newblock Error mitigation for short-depth quantum circuits.
\newblock {\em Physical Review Letters}, 119(18), November 2017.

\bibitem{PhysRevB_scanning_tunneling_micro}
J.~Tersoff and D.~R. Hamann.
\newblock Theory of the scanning tunneling microscope.
\newblock {\em Phys. Rev. B}, 31:805--813, Jan 1985.

\bibitem{Yang_2022}
Bo~Yang, Rudy Raymond, and Shumpei Uno.
\newblock Efficient quantum readout-error mitigation for sparse measurement
  outcomes of near-term quantum devices.
\newblock {\em Physical Review A}, 106(1), July 2022.

\bibitem{zhu2024localization}
Zhaoxuan Zhu, Shengjie Yu, Dean Johnstone, and Laurent Sanchez-Palencia.
\newblock Localization and spectral structure in two-dimensional quasicrystal
  potentials, 2024.

\end{thebibliography}

\end{document}